\catcode`@=11
\global\newcount\secno
\global\newcount\subsecno
\global\newcount\subsubsecno
\global\newcount\equationno
\global\newcount\refno
\global\newcount\footnoteno
\global\newcount\@no
\secno=0\subsecno=0\subsubsecno=0\equationno=0\refno=0\footnoteno=0
\let\\=\cr
\def\@draftleft#1{}
\def\@draftright#1{}
\overfullrule=0pt
\def\draft{\def\@draftleft##1{\leavevmode\vadjust{\smash{%
\raise3pt\llap{\eighttt\string##1~~}}}}%
\def\@draftright##1{\rlap{\eighttt~~\string##1}}%
\def\date##1{\leftline{\number\month/\number\day/\number\year\
\the\time}\bigskip}\overfullrule=5pt\normalbaselineskip=18pt
\normalbaselines}
\def\@the#1{\ifnum\the#1>0\relax\the#1\else\ifnum\the#1<0\relax
\@no=-\the#1\advance\@no'100\char\@no\fi\fi}
\def\@advance#1{\ifnum\the#1<0\global\advance#1 -1\relax
\else\global\advance#1 1\relax\fi}
\def\nsec#1\par{\bigbreak\bigskip\@advance\secno
\subsecno=0\subsubsecno=0\equationno=0
\vbox{\secfont\noindent
\@the\secno. #1\medskip}\nobreak\noindent\ignorespaces}
\def\secadvance{\@advance\secno}
\def\sec#1#2\par{\bigbreak\bigskip
\subsecno=0\subsubsecno=0\equationno=0
\if*#1\vbox{\secfont\noindent\ignorespaces#2\medskip}%
\else
\secno=#1
\vbox{\secfont\noindent\@the\secno. #2\medskip}\fi
\nobreak\noindent\ignorespaces}
\def\seclab#1{\xdef#1{\@the\secno}\@draftleft#1}
\def\nsubsec#1\par{\bigskip\@advance\subsecno
\subsubsecno=0\equationno=0
\vbox{\subsecfont\noindent\@the\secno.\@the\subsecno. #1\medskip}%
\nobreak\noindent\ignorespaces}
\def\subsecadvance{\@advance\subsecno}
\def\subsec#1#2\par{\bigskip
\subsubsecno=0\equationno=0
\if*#1\vbox{\subsecfont\noindent\ignorespaces#2\medskip}%
\else
\subsecno=#1
\vbox{\subsecfont\noindent\@the\secno.\@the\subsecno. #2\medskip}\fi
\nobreak\noindent\ignorespaces}
\def\subseclab#1{\xdef#1{\@the\secno.\@the\subsecno}%
\@draftleft#1}
\def\nsubsubsec#1\par{\medskip\@advance\subsubsecno
\vbox{\subsubsecfont\noindent
\@the\secno.\@the\subsecno.\@the\subsubsecno. #1\medskip}%
\nobreak\noindent\ignorespaces}
\def\subsubsecadvance{\@advance\subsubsecno}
\def\subsubsec#1#2\par{\medskip
\if*#1\vbox{\subsubsecfont\noindent\ignorespaces#2\medskip}%
\else
\subsubsecno=#1
\vbox{\subsubsecfont\noindent
\@the\secno.\@the\subsecno.\@the\subsubsecno. #2\medskip}\fi
\nobreak\noindent\ignorespaces}
\def\subsubseclab#1{\xdef#1{\@the\secno.\@the\subsecno.\@the\subsubsecno}%
\@draftleft#1}
\def\eqlabel#1{\@advance\equationno
\ifnum\secno=0\xdef#1{\the\equationno}\else
\ifnum\subsecno=0\xdef#1{\@the\secno.\the\equationno}\else
\xdef#1{\@the\secno.\@the\subsecno.\the\equationno}\fi\fi
\eqno({\eqnofont #1})\@draftright#1}
\def\lnlabel#1{\global\advance\equationno1
\ifnum\secno=0\xdef#1{\the\equationno}\else
\ifnum\subsecno=0\xdef#1{\@the\secno.\@the\equationno}\else
\xdef#1{\@the\secno.\@the\subsecno.\the\equationno}\fi\fi
&({\eqnofont #1})\@draftright#1}
\def\eqadvance#1{\global\advance\equationno1
\ifnum\secno=0\xdef#1{\the\equationno}\else
\ifnum\subsecno=0\xdef#1{\@the\secno.\the\equationno}\else
\xdef#1{\@the\secno.\@the\subsecno.\the\equationno}\fi\fi}
\def\eqlabelno(#1#2){\eqno({\eqnofont #1#2})\@draftright#1}
\def\lnlabelno(#1#2){&({\eqnofont #1#2})\@draftright#1}
\newwrite\rfile
\def\nref#1#2{\global\advance\refno1\xdef#1{\the\refno}%
\immediate\write
\rfile{\noexpand\item{#1.}\noexpand\@draftleft\noexpand#1%
#2}}
\def\sref#1#2{\immediate\write
\rfile{\noexpand\item{#1.}\noexpand\@draftleft\noexpand#1%
#2}}
\def\refs#1#2 {\if.#2#2$^{\rm#1}$\spacefactor=\sfcode`.{}\space
\else\if,#2#2$^{\rm#1}$\spacefactor=\sfcode`,{}\space
\else\if;#2#2$^{\rm#1}$\spacefactor=\sfcode`;{}\space
\else\if:#2#2$^{\rm#1}$\spacefactor=\sfcode`:{}\space
\else\if?#2#2$^{\rm#1}$\spacefactor=\sfcode`?{}\space
\else\if!#2#2$^{\rm#1}$\spacefactor=\sfcode`!{}\space
\else
$^{\rm#1}$#2\space\fi\fi\fi\fi\fi\fi}
\def\Refs#1{$\rm#1$}
\def\reportno#1{\line{\hfil\vbox{\halign{\strut##\hfil\cr#1\crcr}}}}
\def\Title#1{\vskip3\bigskipamount\line{\titlefont
\hfil\vbox{\halign{\strut\hfil##\hfil\cr#1\crcr}}\hfil}%
\vskip2\bigskipamount}
\def\author#1{\centerline{\authorfont#1}\medskip}
\def\address#1{\centerline{\vbox{\halign
{\strut\hfil\addressfont##\hfil\cr#1\crcr}}}%
\bigskip}
\def\abstract#1{{\narrower\abstractfont\null\bigskip\noindent\ignorespaces
#1\bigskip}}
\def\date#1{\leftline{#1}\bigskip}
\def\Tr{\mathop{\rm Tr}\nolimits}
\immediate\openout\rfile=refs.tmp
\font\seventeenrm=cmr17 \font\fourteenrm=cmr10 scaled 1440
\font\twelverm=cmr12  \font\eightrm=cmr8  \font\sixrm=cmr6
\font\seventeeni=cmmi10 scaled 1728 \font\fourteeni=cmmi10 scaled 1440
\font\twelvei=cmmi12  \font\eighti=cmmi8  \font\sixi=cmmi6
\font\seventeensy=cmsy10 scaled 1728 \font\fourteensy=cmsy10 scaled 1440
\font\twelvesy=cmsy10 scaled 1200 \font\eightsy=cmsy8 \font\sixsy=cmsy6
\font\seventeenbf=cmbx10 scaled 1728 \font\fourteenbf=cmbx10 scaled 1440
\font\twelvebf=cmbx12 \font\eightbf=cmbx8 \font\sixbf=cmbx6
\font\seventeentt=cmtt10 scaled 1728 \font\fourteentt=cmtt10 scaled 1440
\font\twelvett=cmtt12 \font\eighttt=cmtt8
\font\seventeenit=cmti10 scaled 1728 \font\fourteenit=cmti10 scaled 1440
\font\twelveit=cmti12 \font\eightit=cmti8
\font\seventeensl=cmsl10 scaled 1728 \font\fourteensl=cmsl10 scaled 1440
\font\twelvesl=cmsl12 \font\eightsl=cmsl8
\font\seventeenex=cmex10 scaled 1728 \font\fourteenex=cmex10 scaled 1440
\font\twelveex=cmex10 scaled 1200
\def\tenpoint{\def\rm{\fam0\tenrm}%
\textfont0=\tenrm\scriptfont0=\sevenrm\scriptscriptfont0=\fiverm
\textfont1=\teni\scriptfont1=\seveni\scriptscriptfont1=\fivei
\textfont2=\tensy\scriptfont2=\sevensy\scriptscriptfont2=\fivesy
\textfont3=\tenex\scriptfont3=\tenex\scriptscriptfont3=\tenex
\textfont\itfam=\tenit\def\it{\fam\itfam\tenit}%
\textfont\slfam=\tensl\def\sl{\fam\slfam\tensl}%
\textfont\ttfam=\tentt\def\tt{\fam\ttfam\tentt}%
\textfont\bffam=\tenbf\scriptfont\bffam=\sevenbf
\scriptscriptfont\bffam=\fivebf\def\bf{\fam\bffam\tenbf}%
\def\small{\eightpoint}%
\def\large{\twelvepoint}%
\def\normalbaselines{\lineskip\normallineskip
  \baselineskip\normalbaselineskip \lineskiplimit\normallineskiplimit}
\setbox\strutbox=\hbox{\vrule height8.5pt depth3.5pt width0pt}%
\normalbaselines\rm}
\def\twelvepoint{\def\rm{\fam0\twelverm}%
\textfont0=\twelverm\scriptfont0=\eightrm\scriptscriptfont0=\sixrm
\textfont1=\twelvei\scriptfont1=\eighti\scriptscriptfont1=\sixi
\textfont2=\twelvesy\scriptfont2=\eightsy\scriptscriptfont2=\sixsy
\textfont3=\twelveex\scriptfont3=\twelveex\scriptscriptfont3=\twelveex
\textfont\itfam=\twelveit\def\it{\fam\itfam\twelveit}%
\textfont\slfam=\twelvesl\def\sl{\fam\slfam\twelvesl}%
\textfont\ttfam=\twelvett\def\tt{\fam\ttfam\twelvett}%
\textfont\bffam=\twelvebf\scriptfont\bffam=\eightbf
\scriptscriptfont\bffam=\sixbf\def\bf{\fam\bffam\twelvebf}%
\def\small{\tenpoint}%
\def\large{\fourteenpoint}%
\def\normalbaselines{\lineskip1.2\normallineskip
  \baselineskip1.2\normalbaselineskip \lineskiplimit1.2\normallineskiplimit}
\setbox\strutbox=\hbox{\vrule height10.2pt depth4.2pt width0pt}%
\normalbaselines\rm}
\def\fourteenpoint{\def\rm{\fam0\fourteenrm}%
\textfont0=\fourteenrm\scriptfont0=\tenrm\scriptscriptfont0=\sevenrm
\textfont1=\fourteeni\scriptfont1=\teni\scriptscriptfont1=\seveni
\textfont2=\fourteensy\scriptfont2=\tensy\scriptscriptfont2=\sevensy
\textfont3=\fourteenex\scriptfont3=\fourteenex\scriptscriptfont3=\fourteenex
\textfont\itfam=\fourteenit\def\it{\fam\itfam\fourteenit}%
\textfont\slfam=\fourteensl\def\sl{\fam\slfam\fourteensl}%
\textfont\ttfam=\fourteentt\def\tt{\fam\ttfam\fourteentt}%
\textfont\bffam=\fourteenbf\scriptfont\bffam=\tenbf
\scriptscriptfont\bffam=\fivebf\def\bf{\fam\bffam\fourteenbf}%
\def\small{\twelvepoint}%
\def\large{\seventeenpoint}%
\def\normalbaselines{\lineskip1.44\normallineskip
  \baselineskip1.44\normalbaselineskip \lineskiplimit1.44\normallineskiplimit}
\setbox\strutbox=\hbox{\vrule height12.24pt depth5.04pt width0pt}%
\normalbaselines\rm}
\def\seventeenpoint{\def\rm{\fam0\seventeenrm}%
\textfont0=\seventeenrm\scriptfont0=\twelverm\scriptscriptfont0=\eightrm
\textfont1=\seventeeni\scriptfont1=\twelvei\scriptscriptfont1=\eighti
\textfont2=\seventeensy\scriptfont2=\twelvesy\scriptscriptfont2=\eightsy
\textfont3=\seventeenex\scriptfont3=\seventeenex
\scriptscriptfont3=\seventeenex
\textfont\itfam=\seventeenit\def\it{\fam\itfam\seventeenit}%
\textfont\slfam=\seventeensl\def\sl{\fam\slfam\seventeensl}%
\textfont\ttfam=\seventeentt\def\tt{\fam\ttfam\seventeentt}%
\textfont\bffam=\seventeenbf\scriptfont\bffam=\twelvebf
\scriptscriptfont\bffam=\eightbf
\def\bf{\fam\bffam\seventeenbf}%
\def\small{\fourteenpoint}%
\def\large{\seventeenpoint}%
\def\normalbaselines{\lineskip1.73\normallineskip
  \baselineskip1.73\normalbaselineskip \lineskiplimit1.73\normallineskiplimit}
\setbox\strutbox=\hbox{\vrule height14.7pt depth6.0pt width0pt}%
\normalbaselines\rm}
\def\eightpoint{\def\rm{\fam0\eightrm}%
\textfont0=\eightrm\scriptfont0=\sixrm\scriptscriptfont0=\fiverm
\textfont1=\eighti\scriptfont1=\sixi\scriptscriptfont1=\fivei
\textfont2=\eightsy\scriptfont2=\sixsy\scriptscriptfont2=\fivesy
\textfont3=\tenex\scriptfont3=\tenex\scriptscriptfont3=\tenex
\textfont\itfam=\eightit\def\it{\fam\itfam\eightit}%
\textfont\slfam=\eightsl\def\sl{\fam\slfam\eightsl}%
\textfont\ttfam=\eighttt\def\tt{\fam\ttfam\eighttt}%
\textfont\bffam=\eightbf\scriptfont\bffam=\sixbf
\scriptscriptfont\bffam=\fivebf\def\bf{\fam\bffam\eightbf}%
\def\small{\eightpoint}%
\def\large{\tenpoint}%
\def\normalbaselines{\lineskip.8\normallineskip
  \baselineskip.8\normalbaselineskip \lineskiplimit.8\normallineskiplimit}
\setbox\strutbox=\hbox{\vrule height7pt depth3pt width0pt}%
\normalbaselines\rm}
\def\small{\eightpoint}
\def\large{\twelvepoint}
\def\vfootnote#1{\insert\footins\bgroup
  \interlinepenalty\interfootnotelinepenalty
  \splittopskip\ht\strutbox 
  \splitmaxdepth\dp\strutbox \floatingpenalty\@MM
  \leftskip\z@skip \rightskip\z@skip \spaceskip\z@skip \xspaceskip\z@skip
  \footnotefont\textindent{#1}\footstrut\futurelet\next\fo@t}
\def\nfootnote{\advance\footnoteno1\@no=\footnoteno\advance\@no'140
\footnote{$^{\char\@no}$}}
\def\nvfootnote#1{\advance\footnoteno1\@no=\footnoteno\advance\@no'140
\def#1{$^{\char\@no}$}\vfootnote{$^{\char\@no}$}}
\def\titlefont{\seventeenpoint\rm}
\def\authorfont{\twelvepoint\rm}
\def\addressfont{\tenpoint\it}
\def\abstractfont{\eightpoint\rm}
\def\secfont{\tenpoint\bf}
\def\subsecfont{\tenpoint\sl}
\def\subsubsecfont{\tenpoint\it}
\def\eqnofont{\tenpoint\rm}
\def\footnotefont{\eightpoint\rm}
\catcode`@=12

\input pictex
\let\backslash=\\
\let\\=\cr

\def\e{{\rm e}}
\def\i{{\rm i}}
\def\const{\,{\rm const}\,}
\def\Z{{\bf Z}}
\def\C{{\bf C}}

\def\F{{\cal F}}
\def\H{{\cal H}}
\def\P{{\cal P}}
\def\Q{{\cal Q}}
\def\N{{\cal N}}
\def\ve{\varepsilon}
\def\Re{\mathop{\rm Re}\nolimits}
\def\Im{\mathop{\rm Im}\nolimits}
\def\Tr{\mathop{\rm Tr}\nolimits}

\def\Sym{\mathop{\rm Sym}\nolimits}
\def\Res{\mathop{\rm Res}}
\def\half{{1\over2}}
\def\quarter{{1\over4}}
\def\lcolon{\mathopen:}
\def\rcolon{\mathclose:}
\def\W(#1,#2;#3,#4|#5){\mathop W\left[\matrix{#4&#3\cr#1&#2}\bigg|
                        \matrix{#5}\right]}
\def\L(#1,#2;#3,#4|#5){\mathop L\left[\matrix{#4&#3\cr#1&#2}\bigg|
                        \matrix{#5}\right]}
\def\K(#1,#2;#3,#4|#5){\mathop K\left[\matrix{#4&#3\cr#1&#2}\bigg|
                        \matrix{#5}\right]}
\catcode`\@=11
\def\negbigskip{\vskip-\bigskipamount}
\def\ssubsec#1#2\par{\vskip-\lastskip \medskip\bigskip
\subsubsecno=0
\if*#1\vbox{\subsecfont\noindent\ignorespaces#2\medskip}%
\else
\vbox{\subsecfont\noindent\@the\secno.#1. #2\medskip}\fi
\nobreak\noindent\ignorespaces}
\catcode`\@=12
\def\ar#1 #2, #3 #4 {\arrow <0.3truecm> [0.1,0.3] from #1 #2 to #3 #4 }
\catcode`!=11
\def\ln#1 #2, #3 #4 {\!start(#1,#2)\!ljoin(#3,#4)}
\catcode`!=12
\def\rl#1 #2, #3 #4 {\putrule from #1 #2 to #3 #4 }
\def\trl#1 #2, #3 #4 {\linethickness .8pt
	\putrule from #1 #2 to #3 #4 \linethickness .4pt}
\def\point#1 #2 {\put{\hbox{\kern -1pt .}} [Bl] at #1 #2 }
\def\bfpoint#1 #2 {%
\put{\hbox{\kern -2pt \raise -2.3pt%
\hbox{$\textstyle\bullet$}}} [Bl] at #1 #2 }
\def\figcap#1#2\par{\medskip{\narrower\eightpoint\noindent
	Fig.~#1. \ignorespaces#2\par}\medskip}

\reportno{LANDAU-97-TMP-5\\hep-th/9710099}
\Title{Free Field Construction for Correlation Functions\\
of the Eight-Vertex Model}
\author{Michael Lashkevich and Yaroslav Pugai}
\address{L.~D.~Landau Institute for Theoretical Physics,\\
142432 Chernogolovka, Russia}
\abstract{A free field representation for the type $I$
vertex operators and the corner transfer matrices
of the eight-vertex model is proposed.
The construction uses the vertex-face correspondence,
which makes it possible to express
correlation functions of the eight-vertex model
in terms of correlation functions of the SOS model
with a nonlocal insertion. This new nonlocal insertion admits of
a free field representation in terms of Lukyanov's screening
operator. The spectrum of the corner transfer matrix and the Baxter--Kelland
formula for the average staggered polarization have been reproduced.}
\date{October 1997}

\nref\Suth{B.~Sutherland, {\it J.~Math.\ Phys.}\ {\bf 11}, 3183 (1970)}
\nref\FanWu{C.~Fan and F.~Y.~Wu, {\it Phys.\ Rev.}\ {\bf B2}, 723 (1970)}
\nref\Baxter{R.~J.~Baxter, {\it Exactly Solved Models in
	Statistical Mechanics}, Academic Press, 1982}
\nref\BaxterPRL{R.~J.~Baxter, {\it Phys.\ Rev.\ Lett.}\ {\bf 26},
	832 (1971); {\it Ann.\ Phys.\ (N.~Y.)\/} {\bf 70},
	193 (1972)}
\nref\BK{R.~J.~Baxter and S.~B.~Kelland, {\it J.~Phys.}\ {\bf C7}, L403
	(1974)}
\nref\BarBax{M.~N.~Barber and R.~J.~Baxter, {\it J.~Phys.}\ {\bf C6},
	2913 (1973)}
\nref\Baxcor{R.~J.~Baxter, {\it Phil.\ Trans.\ Royal Soc.\ London}\
	{\bf 289}, 315 (1978)}
\nref\JMeightvertex{M.~Jimbo, T.~Miwa, and A.~Nakayashiki,
	{\it J.~Phys.}\ {\bf A26} 2199 (1993)}
\nref\DFJMN{B.~Davies, O.~Foda, M.~Jimbo, T.~Miwa, and A.~Nakayashiki,
	{\it Commun.\ Math.\ Phys.}\ {\bf 151}, 89 (1993)}
\nref\JMbos{M.~Jimbo, K.~Miki, T.~Miwa, and A.~Nakayashiki,
	{\it Phys.\ Lett.}\ {\bf A168}, 256 (1992)}
\nref\JMbook{M.~Jimbo and T.~Miwa, {\it Algebraic Analysis of Solvable
	Lattice Models}, CBMS Regional Conference Series in Mathematics,
	{\bf 85}, AMS, 1994}
\nref\Lukyanov{S.~Lukyanov, {\it Commun.\ Math.\ Phys.}\ {\bf 167},
	183 (1995)}
\nref\Lukjost{S.~Lukyanov, {\it Phys.\ Lett.}\ {\bf B325}, 409 (1994)}
\nref\JMrsos{O.~Foda, M.~Jimbo, T.~Miwa, K.~Miki, and A.~Nakayashiki,
	{\it J.\ Math.\ Phys.}\ {\bf 35}, 13 (1994)}
\nref\LPone{S.~Lukyanov and Ya.~Pugai, {\it J.~Exp.\ Theor.\ Phys.}\
	{\bf 82}, 1021 (1996) ({\tt hep-th/9412128})}
\nref\LP{S.~Lukyanov and Ya.~Pugai, {\it Nucl.\ Phys.}\ {\bf B[FS]473},
	631 (1996) ({\tt hep-th/9602074})}
\nref\AJMP{Y.~Asai, M.~Jimbo, T.~Miwa, and Ya.~Pugai,
	{\it J.~Phys.}\ {\bf A29}, 6595 (1996)}
\nref\FJMOP{B.~Feigin, M.~Jimbo, T.~Miwa, A.~Odesskii, and Ya.~Pugai,
	Algebra of screening operators for the deformed $W_n$ algebra,
	preprint {\tt q-alg/9702029}, (February 1997)}
\nref\Bougourzi{A.~H.~Bougourzi, Bosonization of quantum affine
	groups and its application to the higher spin Heisenberg model,
	preprint ITP-SB-97-29, {\tt q-alg/9706015} (June 1997)}
\nref\Konno{H.~Konno, An elliptic algebra $U_{q,p}(\widehat{sl}_2)$
	and the fusion RSOS model, preprint {\tt q-alg/9709013}
	(September 1997)}
\nref\Martinez{J.~R.~Reyes Martinez, Correlation functions for
	the $\Z$-invariant Ising model, {\tt hep-th/9609135} (September 1996)}
\nref\Bsos{R.~J.~Baxter, {\it Ann.\ Phys.\ (N.~Y.)\/} {\bf 76},
	1, 25, 48 (1973)}
\nref\JMunpublished{M.~Jimbo and T.~Miwa, unpublished, 1993;
	E.~Frenkel, S.~Lukyanov, and N.~Reshetikhin, unpublished, 1996;
	B.~Feigin and A.~Odesskii, unpublished, 1996;
	M.~Lashkevich and Ya.~Pugai, unpublished, 1996;
	H.~Fau, B.~Hou, K.~Shi, and W.~Yang, {\it J.~Phys.}\ {\bf A30},
	5687 (1997)}
\nref\Lashdisorder{M.~Yu.~Lashkevich, {\it Mod.\ Phys.\ Lett.}\ {\bf B10},
	101 (1996) ({\tt hep-th/9408131})}
\nref\KashMiwa{S.-J.~Kang, M.~Kashiwara, K.~C.~Misra, T.~Miwa,
	T.~Nakashima, and A.~Nakayashiki, {\it C.~R.\ Acad.\ Sci.\ Paris},
	{\bf 315I}, 375 (1992)}
\nref\JMellipticalg{O.~Foda, K.~Iohara, M.~Jimbo, R.~Kedem, T.~Miwa,
	and H.~Yan, {\it Lett.\ Math.\ Phys.}\ {\bf 32},
	258 (1994)}
\nref\JLMP{M.~Jimbo, M.~Lashkevich, T.~Miwa, Ya.~Pugai,
	{\it Phys.\ Lett.}\ {\bf A229}, 285 (1997) ({\tt hep-th/9607177})}

\nsec Introduction

The eight-vertex model\refs{\Suth,\FanWu,\Baxter} is one of the most
famous examples of exactly solvable models of statistical mechanics.
Its partition function was found in 1971 by Baxter\refs{\BaxterPRL}.
Later expressions for the spontaneous polarization and magnetization were
proposed\refs{\BK,\BarBax}. Nevertheless, very few results on correlation
functions have been obtained up to now\refs{\Baxter,\Baxcor,\JMeightvertex}.

For the last years, a new approach to calculating correlation
functions of integrable (exactly solvable) lattice and continuous
models has been developed\refs{\DFJMN-\Lukjost}.
This approach, known as the vertex operator approach, has shown its
efficiency in
calculation and analysis of correlation functions and form factors
for different integrable models,
such as the six-vertex
model\refs{\DFJMN-\JMbook},
the restricted solid-on-solid (RSOS) models\refs{\JMrsos-\LP}
and their $A^{(1)}_n$ generalizations\refs{\AJMP,\FJMOP},
the fusion RSOS models\refs{\Bougourzi,\Konno},
the Ising model\refs{\JMrsos,\Martinez}, the sine-Gordon model\refs{\Lukyanov,%
\Lukjost}.
The most important constituent of this approach that allows one to perform
calculations is the free field representation of the vertex operators.

However, the free field representation was not elaborated for the
eight-vertex model. Formally, the problem lies in the presence
of the $d$ matrix element in the $R$ matrix of this model which breaks down
the charge conservation, which is necessary for the
free field representation. In principle, the same reason gives no way
of applying directly the Bethe ansatz method
to this model\refs{\Baxter}. Baxter's idea to overcome this difficulty
for the Bethe ansatz
was to twist somehow the model to restore the charge conservation law
without change of eigenvalues of the transfer matrix\refs{\Bsos}.
Such `twisted'
model was the solid-on-solid (SOS) model, which is a face type model.
It is related to the eight-vertex model by the so-called vertex-face
correspondence.

It is an old idea to use this vertex-face correspondence to
relate correlation functions of the eight-vertex model with ones of
the SOS model\refs{\JMunpublished},
and then to apply the known free field representation
for the vertex operators of the SOS model.
The obstacle is non-local
character of the vertex-face correspondence for correlation functions.
In this paper we analyze this relation and propose a free field approach
for the correlation functions of the eight-vertex model. As a check we
find the spectrum of the corner transfer matrix of the eight-vertex model,
and reproduce the Baxter--Kelland formula\refs{\BK} for the staggered
spontaneous polarization in the antiferroelectric phase.

The paper is organized as follows. In Sec.~2 basic definitions
and notations are given. In Sec.~3 we give a review
of the vertex operator approach of both the eight-vertex and the SOS
models. In Sec.~4 we use the vertex-face correspondence to find a relation
between correlation function of the eight-vertex model and the SOS model.
The construction contains a new object in addition
to corner transfer matrices and vertex operators. This object
(operator $\Lambda$ in our designations) relates the corner
transfer matrices of the eight-vertex model to those of
the SOS model. In Sec.~5 we describe a free field
representation and calculate the
staggered polarization by use of the free field representation.
In Sec.~6 we discuss main conjectures used
in the paper and possible directions of future studies.

\nsec Basic Definitions and Vertex-Face Correspondence\negbigskip

\ssubsec{1}Eight-Vertex Model

The eight-vertex model is defined as follows\refs{\Baxter}.
The fluctuating variables (polarizations)
$\ve=\pm1\equiv\pm$ are placed at links of a square lattice.
Interaction is associated with vertices of the lattice. A local weight
$R_{\ve_1\ve_2}^{\ve_3\ve_4}$ is associated to each configuration
of polarizations $\ve_1$, $\ve_2$, $\ve_3$, $\ve_4$
around a vertex (Fig.~1a).
A weight of a lattice configuration is the product of local weights.

It is convenient to attach a constant
complex `spectral parameter' to each
line of the lattice, and to consider a weight as a function of
the difference of the spectral parameters of the two lines that intersect at
the vertex. The weights as functions of this difference
are given by\refs{\Baxter}
$$
\eqalign{
a(u)=R(u)_{++}^{++}=R(u)_{--}^{--}
&=-\i\kappa (u) R_0(u)\,
\textstyle
\theta_4\left(\i{\epsilon\over\pi};\i{2\epsilon r\over\pi}\right)
\theta_4\left(\i{\epsilon\over\pi}u;\i{2\epsilon r\over\pi}\right)
\theta_1\left(\i{\epsilon\over\pi}(1-u);\i{2\epsilon r\over\pi}\right),
\cr
b(u)=R(u)_{+-}^{+-}=R(u)_{-+}^{-+}
&=-\i\kappa (u) R_0(u)\,
\textstyle
\theta_4\left(\i{\epsilon\over\pi};\i{2\epsilon r\over\pi}\right)
\theta_1\left(\i{\epsilon\over\pi}u;\i{2\epsilon r\over\pi}\right)
\theta_4\left(\i{\epsilon\over\pi}(1-u);\i{2\epsilon r\over\pi}\right),
\cr
c(u)=R(u)_{+-}^{-+}=R(u)_{-+}^{+-}
&=-\i\kappa (u) R_0(u)\,
\textstyle
\theta_1\left(\i{\epsilon\over\pi};\i{2\epsilon r\over\pi}\right)
\theta_4\left(\i{\epsilon\over\pi}u;\i{2\epsilon r\over\pi}\right)
\theta_4\left(\i{\epsilon\over\pi}(1-u);\i{2\epsilon r\over\pi}\right),
\cr
d(u)=R(u)_{++}^{--}=R(u)_{--}^{++}
&=-\i\kappa (u) R_0(u)\,
\textstyle
\theta_1\left(\i{\epsilon\over\pi};\i{2\epsilon r\over\pi}\right)
\theta_1\left(\i{\epsilon\over\pi}u;\i{2\epsilon r\over\pi}\right)
\theta_1\left(\i{\epsilon\over\pi}(1-u);\i{2\epsilon r\over\pi}\right)
}\eqadvance\EQRmatrix
\eqlabelno(\EQRmatrix a)
$$
with $\theta_i(u;\tau)$ is the $i$th theta function with the basic periods
$1$ and $\tau$ ($\Im\tau>0$). Here $\epsilon$ and $r$ are parameters.
It is often convenient to use another set of parameters
$$
x=\e^{-\epsilon},\qquad p=x^{2r},\qquad z=x^{2u}.
\eqlabel\EQparameters
$$
We shall use both `additive' parameters $u$, $\epsilon$, $r$ and
`multiplicative' parameters $z$, $x$, $p$ on equal grounds.

\topinsert
%
%
\line{\hfil
\beginpicture
\setcoordinatesystem units <1cm,1cm> point at 0 -0.5
\rl 2.5 2 , 2.5 0.8 \ar 2.5 0.81 , 2.5 0.8
\rl 3 1.5 , 1.8 1.5 \ar 1.81 1.5 , 1.8 1.5
\put{$R(u-v)_{\ve_1\ve_2}^{\ve_3\ve_4}=$} [Br] at 1.2 1.4
\put{$\ve_1$} [Bl] at 2.6 0.9
\put{$\ve_2$} [Bl] at 1.9 1.7
\put{$\ve_3$} [Bl] at 2.6 1.9
\put{$\ve_4$} [Bl] at 3.0 1.6
\put{$u$} [t] at 2.5 0.6
\put{$v$} [rB] at 1.7 1.5
\put{$(a)$} [Bl]  at 1.6 -0.7
\setcoordinatesystem units <1cm,1cm> point at -4 0
\rl 1.0 3.3 , 1.0 0.7
\rl 2.0 3.3 , 2.0 0.7
\rl 3.0 3.3 , 3.0 0.7
\rl 0.7 3.0 , 3.3 3.0
\rl 0.7 2.0 , 3.3 2.0
\rl 0.7 1.0 , 3.3 1.0
\put{$+$} [lb] at 1.1 3.4
\put{$+$} [lb] at 0.4 3.1
\put{$+$} [lb] at 3.1 3.4
\put{$+$} [lb] at 2.4 3.1
\put{$+$} [lb] at 2.1 2.4
\put{$+$} [lb] at 1.4 2.1
\put{$+$} [lb] at 1.1 1.4
\put{$+$} [lb] at 0.4 1.1
\put{$+$} [lb] at 3.4 2.1
\put{$+$} [lb] at 3.1 1.4
\put{$+$} [lb] at 2.4 1.1
\put{$+$} [lb] at 2.1 0.4
\put{$-$} [lb] at 2.1 3.4
\put{$-$} [lb] at 1.4 3.1
\put{$-$} [lb] at 1.1 2.4
\put{$-$} [lb] at 0.4 2.1
\put{$-$} [lb] at 3.4 3.1
\put{$-$} [lb] at 3.1 2.4
\put{$-$} [lb] at 2.4 2.1
\put{$-$} [lb] at 2.1 1.4
\put{$-$} [lb] at 1.4 1.1
\put{$-$} [lb] at 1.1 0.4
\put{$-$} [lb] at 3.4 1.1
\put{$-$} [lb] at 3.1 0.4
\put{$(b)$} [B] at 4.0 -0.2
\setcoordinatesystem units <1cm,1cm> point at -8 0
\rl 1.0 3.3 , 1.0 0.7
\rl 2.0 3.3 , 2.0 0.7
\rl 3.0 3.3 , 3.0 0.7
\rl 0.7 3.0 , 3.3 3.0
\rl 0.7 2.0 , 3.3 2.0
\rl 0.7 1.0 , 3.3 1.0
\put{$-$} [lb] at 1.1 3.4
\put{$-$} [lb] at 0.4 3.1
\put{$-$} [lb] at 3.1 3.4
\put{$-$} [lb] at 2.4 3.1
\put{$-$} [lb] at 2.1 2.4
\put{$-$} [lb] at 1.4 2.1
\put{$-$} [lb] at 1.1 1.4
\put{$-$} [lb] at 0.4 1.1
\put{$-$} [lb] at 3.4 2.1
\put{$-$} [lb] at 3.1 1.4
\put{$-$} [lb] at 2.4 1.1
\put{$-$} [lb] at 2.1 0.4
\put{$+$} [lb] at 2.1 3.4
\put{$+$} [lb] at 1.4 3.1
\put{$+$} [lb] at 1.1 2.4
\put{$+$} [lb] at 0.4 2.1
\put{$+$} [lb] at 3.4 3.1
\put{$+$} [lb] at 3.1 2.4
\put{$+$} [lb] at 2.4 2.1
\put{$+$} [lb] at 2.1 1.4
\put{$+$} [lb] at 1.4 1.1
\put{$+$} [lb] at 1.1 0.4
\put{$+$} [lb] at 3.4 1.1
\put{$+$} [lb] at 3.1 0.4
\endpicture
\hfil}
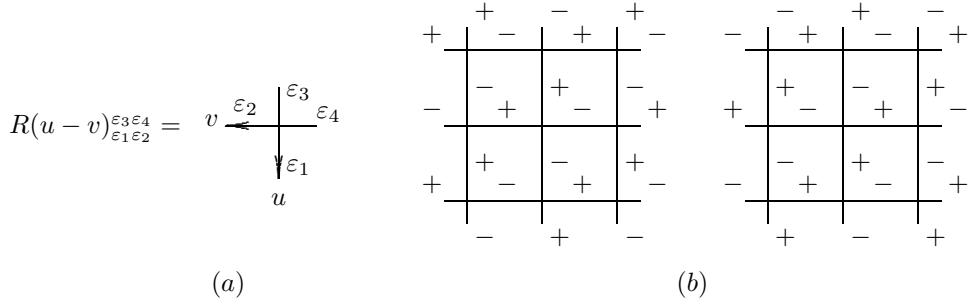
\figcap{1}Eight-vertex model: $(a)$ definition of the weight matrix;
$(b)$ two degenerate ground states.\par
\endinsert

The common factor $\kappa (u)R_0(u)$ is chosen so that
the partition function per site is equal to one:
$$
\eqalignno{
\kappa (u)
&=z^{-{r-1\over 2r}}x^{1-{r\over2}}
(x^{2r};x^{2r})_\infty^{-2}(x^{4r};x^{4r})_\infty^{-1}
(x^2z^{-1};x^{2r})_\infty^{-1}(x^{2r-2}z;x^{2r})_\infty^{-1},
\lnlabelno(\EQRmatrix b)
\cr
R_0(u)&
=z^{r-1\over2r}
{(x^2z^{-1};x^4,x^{2r})_\infty(x^{2r+2}z^{-1};x^4,x^{2r})_\infty
(x^4z;x^4,x^{2r})_\infty(x^{2r}z;x^4,x^{2r})_\infty
\over
(x^4z^{-1};x^4,x^{2r})_\infty(x^{2r}z^{-1};x^4,x^{2r})_\infty
(x^2z;x^4,x^{2r})_\infty(x^{2r+2}z;x^4,x^{2r})_\infty},
\lnlabelno(\EQRmatrix c)
\cr
\span
(z;p_1,\ldots,p_N)
=\prod_{n_1,\ldots,n_N=0}^\infty(1-zp_1^{n_1}\ldots p_N^{n_N}).
}
$$
Splitting the common factor into two parts is convenient for later use.

The matrix $R(u)$ satisfies the Yang--Baxter equation, the crossing
and unitarity properties\refs{\Baxter,\JMeightvertex}. This supplies
commutativity of transfer matrices with different spectral parameters $u$
(but coincident $\epsilon$ and $r$) and integrability of the model.

For real parameters $\epsilon$, $r$, $u$ in the region
$$
\epsilon>0,\qquad r>1,\qquad -1<u<1
\quad\hbox{or}\quad a+|b|+|d|<c
\eqlabel\EQantiferregion
$$
the model is in the antiferroelectric phase. In the `low temperature' limit
$\epsilon\to\infty$ (or $a,|b|,|d|\ll c$) the system falls in
one of two ground states (Fig.~1b).
These two ground states correspond to the broken $\Z_2$ symmetry.
At finite temperatures the symmetry remains broken in the whole
region (\EQantiferregion). The second order phase transition
takes place at $\epsilon=0$.
The weight $d_A$ in our definition is negative but it
enters into any configuration in an even power in
the antiferroelectric phase and its sign is inessential.

The aim of the present paper is to find a procedure for calculating
correlation functions in the eight-vertex model
in the thermodynamic limit. In our normalization
the partition function per site is equal to 1. So we can define
the correlation function directly on the infinite lattice. Let us
label the vertices of the lattice of finite size by two integer Cartesian
coordinates $(x,y)$. The coordinates of polarizations $\ve(x,y)$
are the coordinates
of the centers of links ($x\in\Z+\half$, $y\in\Z$ for vertical links and
$x\in\Z$, $y\in\Z+\half$ for horizontal ones).
The probability that polarizations at definite links $(x_k,y_k)$,
$k=1,\ldots,N$, take fixed values $\ve_k$ is given by
$$
P^{(i)}_{\ve_1\ldots\ve_N}
=\sum_{\{\ve(x,y)\}\atop\ve(x_k,y_k)=\ve_k}
\prod_{x,y\in\Z}
R(u_x-v_y)_{\ve(x,y-\half)\,\ve(x-\half,y)}^{\ve(x,y+\half)\,\ve(x+\half,y)}.
\eqlabel\EQprobdef
$$
Here the sum is taken over all configurations such that
$\ve_{x_k,y_k}=\ve_k$ and $\ve_{x,y}$ only differ from the ground
state configuration value $(-)^{x-y-\half+i}$ at a finite number
of links. Note, that we consider an inhomogeneous lattice,
where each line brings its own spectral parameter $u_x$ of $v_y$,
so that $u_x=u=\const$ and $v_y=v=\const$ for large enough $x$ and $y$
respectively.
Later we concentrate our attention on the particular
case of $N$ parallel neighboring links
($x_k=\half$, $y_k=k$, see Fig.~5a), but our consideration
is valid in general case.

\ssubsec{2}SOS model

Let us turn to the SOS model. Here one associates an integer
variable (`height') $n$ to each site of a square lattice, so that the
heights $n_1$ and $n_2$ at adjacent sites satisfy the
admissibility condition
$$
|n_1-n_2|=1.
\eqlabel\EQadmissibility
$$
Interaction is described by weights $\W(n_1,n_2;n_3,n_4|u)$
associated to faces, and the spectral parameters
are attached to lines of the dual lattice (Fig.~2):
$$
\eqadvance\EQWmatrix
\eqalignno{
\W(n\pm1,n\pm2;n\pm1,n|u)
&=R_0(u),
\lnlabelno(\EQWmatrix a)
\cr
\W(n\pm1,n;n\pm1,n|u)
&=R_0(u){[n\pm u][1]\over[n][1-u]},
\lnlabelno(\EQWmatrix b)
\cr
\W(n\mp1,n;n\pm1,n|u)
&=-R_0(u){[n\pm1][u]\over[n][1-u]}.
\lnlabelno(\EQWmatrix c)}
$$
Here we use the designation
$$
\eqalignno{
\span
[u]=x^{{u^2\over r}-u}\Theta_{x^{2r}}(x^{2u})
=\sqrt{\pi\over\epsilon r}\,\e^{\quarter\epsilon r}\,
\theta_1\!\left({u\over r};{\i\pi\over\epsilon r}\right),
\cr
\span
\Theta_p(z)=(z;p)_\infty(pz^{-1};p)_\infty(p;p)_\infty.
}
$$
Note that the weights $W$ contain $[n]$ in the denominators. So the
construction demands some regularization. Namely, we may think that
the heights $n$ belong to the set $\Z+\delta$ with some
real $\delta$. Such shift does not break down the admissibility
condition (\EQadmissibility). After calculation of all physical quantities
one may take the limit $\delta\to0$.
The consideration below is valid literally for the
regularized case, and we shall imply this regularization everywhere.
Moreover, the final results do not depend on $\delta$ at all.

\topinsert
%
%
\line{\hfil
\beginpicture
\setcoordinatesystem units <1cm,1cm> point at 0 0
\put{$\W(n_1,n_2;n_3,n_4|u-v)=$} [rB] at 3.7 2.4
\rl 5.0 2.0 , 6.0 2.0
\rl 6.0 2.0 , 6.0 3.0
\rl 6.0 3.0 , 5.0 3.0
\rl 5.0 3.0 , 5.0 2.0
\setdashes <2.5pt>
\rl 5.5 3.3 , 5.5 1.5
\rl 6.3 2.5 , 4.5 2.5
\setsolid
\ar 5.5 1.51 , 5.5 1.5
\ar 4.51 2.5 , 4.5 2.5
\put{$n_1$} [tr] at 4.9 1.9
\put{$n_2$} [tl] at 6.1 1.9
\put{$n_3$} [bl] at 6.1 3.1
\put{$n_4$} [br] at 4.9 3.1
\put{$u$} [t]  at 5.5 1.4
\put{$v$} [rB] at 4.4 2.5
\put{$(a)$} [B] at 3.5 0.0
\setcoordinatesystem units <1cm,1cm> point at -7 0
\rl 1.0 1.0 , 4.0 1.0
\rl 1.0 2.5 , 4.0 2.5
\rl 1.0 4.0 , 4.0 4.0
\rl 1.0 1.0 , 1.0 4.0
\rl 2.5 1.0 , 2.5 4.0
\rl 4.0 1.0 , 4.0 4.0
\put{$m+s$} [lb] at 1.1 4.1
\put{$m+s$} [lb] at 4.1 4.1
\put{$m+s$} [lb] at 2.6 2.6
\put{$m+s$} [lb] at 1.1 1.1
\put{$m+s$} [lb] at 4.1 1.1
\put{$m$} [lb] at 2.6 4.1
\put{$m$} [lb] at 1.1 2.6
\put{$m$} [lb] at 4.1 2.6
\put{$m$} [lb] at 2.6 1.1
\put{$(b)$} [B] at 2.5 0.0
\endpicture
\hfil}
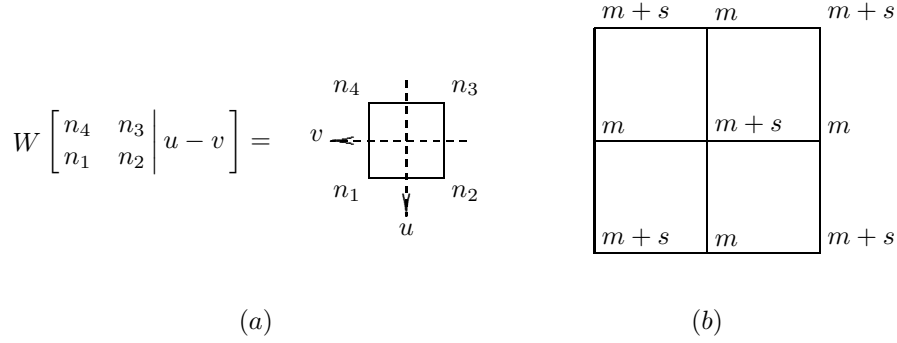
\figcap{2} SOS model: $(a)$ definition of weights; $(b)$ an infinite number
of degenerate ground states labeled by an integer $m$
and by $s=\pm1$ satisfying $(k-1)r<m,m+s<kr$ for some integer $k$.\par
\endinsert

The weights (\EQWmatrix) satisfy a face version of the Yang--Baxter equation,
the crossing and unitarity pro\-per\-ties\refs{\Bsos,\JMrsos}. Note that
the elements (\EQWmatrix a) are analogues of the element $a(u)$ of the
eight-vertex model, (\EQWmatrix b) are analogues of $c(u)$, and (\EQWmatrix c)
of $b(u)$. Namely, we can associate a sign variable $\ve$ to
each link of the dual lattice intersecting a link of the direct
lattice between $n+\ve$ and $n$ ($n+\ve$ is to the left or upper than $n$).
It is impossible to imagine any analog of $d(u)$ in this
picture. The charge defined as
the sum of the variables $\ve$ along an infinite line consisting
of links of the direct lattice is conserved by definition.

We are interested in the so called regime $III$:
$$
\epsilon>0,\qquad r\geq1, \qquad 0<u<1.
\eqlabel\EQregimeIII
$$
There is an infinite number of
degenerate, but inequivalent, ground states
in this regime shown in Fig.~2b.

\ssubsec{3}Vertex-Face Correspondence

Consider the functions (Fig.~3a)
$$
t_\ve(u)^{n'}_n
={\ve^{n-m}\over\sqrt2}\left(
\theta_2\!\left({(n'-n)u+n'\over r};\i{2\pi\over\epsilon r}\right)+
\ve\theta_3\!\left({(n'-n)u+n'\over r};\i{2\pi\over\epsilon r}\right)
\right)
\eqlabel\EQintertwining
$$
with $m$ being an arbitrary integer (in the regularized version
$m\in\Z+\delta$), $\ve=\pm$ being a polarization
variable, $n$ and $n'$ being an admissible pair of heights ({\it i.~e.}\
$|n'-n|=1$).
These functions enter into the identity\refs{\Bsos} (Fig.~4a)
$$
\sum_{\ve'_1\ve'_2}R(u-v)_{\ve_1\ve_2}^{\ve'_1\ve'_2}
t_{\ve'_1}(u_0-u)^{n'}_{s'} t_{\ve'_2}(u_0-v)^{s'}_n
=\sum_{s\in\Z} t_{\ve_2}(u_0-v)^{n'}_s t_{\ve_1}(u_0-u)^s_n
\W(s,n;s',n'|u-v),
\eqlabel\EQvertexface
$$
which is referred to as vertex-face correspondence. The functions
$t_\ve(u)^{n'}_n$ are called intertwining vectors (the subscript $\ve$
is often regarded as a vector index). Note that the identity
holds for arbitrary value of $u_0$.
The vertex-face correspondence is the basic
relation for solving the eight-vertex model by Bethe ansatz. Below we
apply it to find the free field representation. The intertwining vectors
may be pulled through the lattice by use of the vertex-face correspondence.%
\nfootnote{The standard vertex-face correspondence,
as it is described in Ref.~\Refs{\Bsos}, relates the SOS model
in the regime $III$ to
the eight-vertex model in the disordered region. But we can
relate the disordered and ferroelectric region by
use of the Baxter duality transformation\refs{\Baxter}
$$
\eqalign{
a(u)
&={\textstyle\half}(a_D(u)+b_D(u)+c_D(u)+d_D(u)),
\cr
b(u)
&={\textstyle\half}(-a_D(u)-b_D(u)+c_D(u)+c_D(u)),
\cr
c(u)
&={\textstyle\half}(a_D(u)-b_D(u)+c_D(u)-d_D(u)),
\cr
d(u)
&={\textstyle\half}(-a_D(u)+b_D(u)+c_D(u)-d_D(u)),
}
$$
where $a_D$, $b_D$, $c_D$, $d_D$ mean the weights in the
disordered phase. This duality for correlation functions\refs{\Lashdisorder}
gives
$$
t_\ve(u)^{n'}_n={\ve^{n-m}\over\sqrt2}(t^D_+(u)^{n'}_n+\ve t^D_-(u)^{n'}_n).
$$}

\topinsert
%
%
\line{\hfil
\beginpicture
\setcoordinatesystem units <1cm,1cm> point at 0 0
\put{$t_\ve(u_0-u)^{n'}_n=$} [rB] at 3.2 4.4
\trl 4.0 4.5 , 5.0 4.5
\ln 4.4 4.5 , 4.5 4.4
\ln 4.5 4.4 , 4.6 4.5
\rl 4.5 4.4 , 4.5 3.3 \ar 4.5 3.31 , 4.5 3.3
\setdashes <2.5pt>
\rl 3.8 4.2 , 5.2 4.2
\setsolid
\ar 3.81 4.2 , 3.8 4.2
\put{$n$} [b] at 5.0 4.6
\put{$n'$} [b] at 4.0 4.6
\put{$\ve$} [lB] at 4.6 3.8
\put{$u$} [t] at 4.5 3.2
\put{$u\smash{{}_0}$} [r] at 3.7 4.2
\put{$t^*_\ve(u_0-u)^{n'}_n=$} [rB] at 3.2 1.5
\trl 4.0 1.5 , 5.0 1.5
\ln 4.4 1.5 , 4.5 1.6
\ln 4.5 1.6 , 4.6 1.5
\rl 4.5 1.6 , 4.5 2.4
\setdashes <2.5pt>
\rl 4.5 0.9 , 4.5 1.5
\rl 3.8 1.8 , 5.2 1.8
\setsolid
\ar 4.5 0.91 , 4.5 0.9
\ar 3.81 1.8 , 3.8 1.8
\put{$n$} [rt] at 4.0 1.4
\put{$n\smash{{}'}$} [lt] at 5.0 1.4
\put{$\ve$} [lB] at 4.6 2.0
\put{$u$} [t] at 4.5 0.8
\put{$u\smash{{}_0}$} [r] at 3.7 1.8
\put{$(a)$} [B] at 3.7 0.0
\setcoordinatesystem units <1cm,1cm> point at -7 0
\trl 1.0 4.5 , 2.0 4.5
\ln 1.4 4.5 , 1.5 4.4 \ln 1.5 4.4 , 1.6 4.5
\trl 1.0 3.5 , 2.0 3.5
\ln 1.4 3.5 , 1.5 3.6 \ln 1.5 3.6 , 1.6 3.5
\rl 1.5 4.4 , 1.5 3.6
\setdots <2.5pt>
\rl 1.0 4.5 , 1.0 3.5
\setsolid
\put{$n$} [rB] at 0.9 4.6
\put{$n$} [rt] at 0.9 3.4
\put{$n\smash{{}'}$} [lt] at 2.1 3.4
\put{$n''$} [lB] at 2.1 4.6
\put{$t$}   [B] at 1.5 4.6
\put{$t^*$} [t] at 1.5 3.4
\put{$=\delta_{n'n''}$} [lB] at 2.9 4.0
\trl 1.0 2.5 , 2.0 2.5
\ln 1.4 2.5 , 1.5 2.4 \ln 1.5 2.4 , 1.6 2.5
\linethickness=1pt
\rl 1.41 2.49 , 1.59 2.49 \rl 1.44 2.46 , 1.56 2.46
\rl 1.43 2.47 , 1.57 2.47 \rl 1.47 2.43 , 1.53 2.43
\linethickness=.4pt
\trl 1.0 1.5 , 2.0 1.5
\ln 1.4 1.5 , 1.5 1.6 \ln 1.5 1.6 , 1.6 1.5
\rl 1.5 2.4 , 1.5 1.6
\setdots <2.5pt>
\rl 2.0 2.5 , 2.0 1.5
\setsolid
\put{$n''$} [rB] at 0.95 2.6
\put{$n\smash{{}'}$} [rt] at 0.95 1.4
\put{$n$} [lt] at 2.1 1.4
\put{$n$} [lB] at 2.1 2.6
\put{$t'$}  [B] at 1.5 2.6
\put{$t^*$} [t] at 1.5 1.4
\put{$=\delta_{n'n''}$} [lB] at 2.9 2.0
\put{$(b)$} [B] at 2.5 0.0
\endpicture
\hfil}
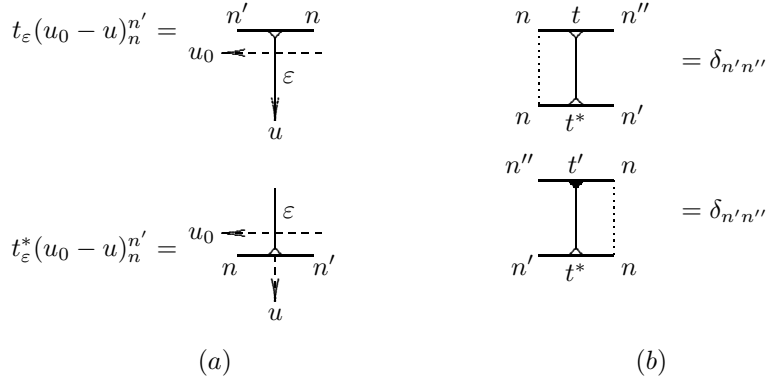
\figcap{3} Intertwining vectors: $(a)$ graphic notations; $(b)$ definition
of the conjugate and `primed' vectors.\par
\endinsert

\topinsert
%
%
\line{\hfil
\beginpicture
\setcoordinatesystem units <1cm,1cm> point at 0 0
\trl 1.0 2.5 , 2.0 2.5
\ln 1.4 2.5 , 1.5 2.4 \ln 1.5 2.4 , 1.6 2.5
\trl 2.0 2.5 , 2.0 1.5
\ln 2.0 2.1 , 1.9 2.0 \ln 1.9 2.0 , 2.0 1.9
\rl 1.5 2.4 , 1.5 1.1 \ar 1.5 1.11 , 1.5 1.1
\rl 1.9 2.0 , 0.6 2.0 \ar 0.61 2.0 , 0.6 2.0
\put{$n$}  [lt] at 2.1 1.4
\put{$s'$} [lB] at 2.1 2.6
\put{$n'$} [rB] at 0.9 2.6
\put{$\ve_1$} [lB] at 1.6 1.4
\put{$\ve_2$} [lt] at 1.0 1.9
\put{$u$} [t]  at 1.5 1.0
\put{$v$} [rB] at 0.5 1.95
\put{$=$} [B] at 2.75 1.95
\trl 4.0 2.5 , 4.0 1.5
\ln 4.0 2.1 , 3.9 2.0 \ln 3.9 2.0 , 4.0 1.9
\trl 4.0 1.5 , 5.0 1.5
\ln 4.4 1.5 , 4.5 1.4 \ln 4.5 1.4 , 4.6 1.5
\rl 4.0 2.5 , 5.0 2.5
\rl 5.0 2.5 , 5.0 1.5
\rl 3.9 2.0 , 3.4 2.0 \ar 3.41 2.0 , 3.4 2.0
\rl 4.5 1.4 , 4.5 0.9 \ar 4.5 0.91 , 4.5 0.9
\setdashes <2.5pt>
\rl 5.3 2.0 , 4.0 2.0
\rl 4.5 2.8 , 4.5 1.5
\setsolid
\put{$n$}  [lt] at 5.1 1.4
\put{$s'$} [lB] at 5.1 2.6
\put{$n'$} [rB] at 3.9 2.6
\put{$\ve_1$} [lB] at 4.65 1.1
\put{$\ve_2$} [lt] at 3.6 1.85
\put{$u$} [t]  at 4.5 0.8
\put{$v$} [rB] at 3.3 1.95
\put{$(a)$} [B] at 3.0 0.0
\setcoordinatesystem units <1cm,1cm> point at -7 0
\trl 1.0 2.5 , 1.0 1.5
\ln 1.0 2.1 , 1.1 2.0 \ln 1.1 2.0 , 1.0 1.9
\trl 1.0 1.5 , 2.0 1.5
\ln 1.4 1.5 , 1.5 1.6 \ln 1.5 1.6 , 1.6 1.5
\rl 1.1 2.0 , 2.0 2.0
\rl 1.5 1.6 , 1.5 2.5
\setdashes <2.5pt>
\rl 1.0 2.0 , 0.5 2.0
\rl 1.5 1.5 , 1.5 1.0
\setsolid
\ar 0.51 2.0 , 0.5 2.0
\ar 1.5 1.01 , 1.5 1.0
\put{$n$}  [lt] at 2.1 1.4
\put{$s$}  [rt] at 0.9 1.4
\put{$n'$} [rB] at 1.0 2.6
\put{$\ve'_1$} [lB] at 1.6 2.5
\put{$\ve'_2$} [lB] at 2.0 2.1
\put{$u$} [t]  at 1.5 0.9
\put{$v$} [rB] at 0.4 1.95
\put{$=$} [B] at 2.75 1.95
\trl 5.0 1.5 , 5.0 2.5
\ln 5.0 1.9 , 5.1 2.0 \ln 5.1 2.0 , 5.0 2.1
\trl 5.0 2.5 , 4.0 2.5
\ln 4.6 2.5 , 4.5 2.6 \ln 4.5 2.6 , 4.4 2.5
\rl 5.0 1.5 , 4.0 1.5
\rl 4.0 1.5 , 4.0 2.5
\rl 5.1 2.0 , 5.3 2.0
\rl 4.5 2.6 , 4.5 2.8
\setdashes <2.5pt>
\rl 5.0 2.0 , 3.4 2.0
\rl 4.5 2.5 , 4.5 1.0
\setsolid
\ar 3.41 2.0 , 3.4 2.0
\ar 4.5 1.01 , 4.5 1.0
\put{$n$}  [lt] at 5.1 1.4
\put{$s$}  [rt] at 3.9 1.4
\put{$n'$} [rB] at 3.9 2.6
\put{$\ve'_1$} [lB] at 4.5 2.8
\put{$\ve'_2$} [lB] at 5.3 2.1
\put{$u$}  [t] at 4.5 0.9
\put{$v$} [rB] at 3.3 1.95
\put{$(b)$} [B] at 3.0 0.0
\endpicture
\hfil}
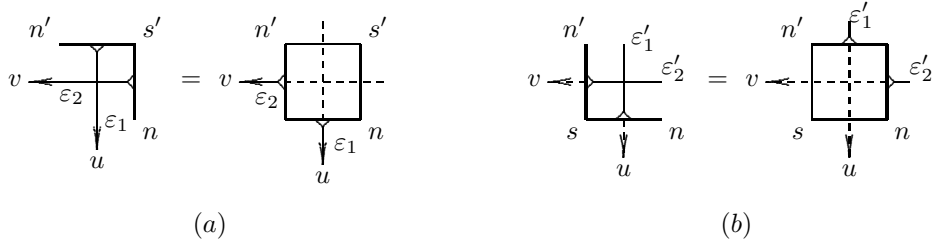
\figcap{4} Vertex-face correspondence: $(a)$ by usual intertwining vectors;
$(b)$ by conjugate intertwining vectors.\par
\endinsert

It is convenient to introduce a `conjugate' and a `primed' intertwining
vectors (Fig.~3b)
$$
\sum_\ve t^*_\ve(u)^{n'}_n t_\ve(u)_{n''}^n = \delta_{n'n''},
\qquad
\sum_\ve t^*_\ve(u)^n_{n'} t'_\ve(u)_n^{n''} = \delta_{n'n''}.
\eqlabel\EQstarprimedef
$$
Explicitly,
$$
\eqalign{
t^*_\ve(u)_n^{n'}
&=(-)^{n-m+1}{n'-n\over C[n][u]}\,t_{-\ve}(u-1)^{n'}_n,
\qquad C=2\theta_2(0)\theta_3(0)[r/2]^{-2},
\cr
t'_\ve(u)_n^{n'}
&={[u]\over[u-1]}{[n']\over[n]}\,t_\ve(u-2)_n^{n'}.
}\eqlabel\EQstarprimeexplicit
$$
The conjugate intertwining vectors enter into another form of
the vertex-face correspondence (Fig.~4b):
$$
\sum_{\ve_1\ve_2}
t^*_{\ve_2}(u_0-v)_{n'}^s t^*_{\ve_1}(u_0-u)_s^n
R(u-v)^{\ve'_1\ve'_2}_{\ve_1\ve_2}
=\sum_{s'\in\Z}
\W(s,n;s',n'|u-v)
t^*_{\ve'_1}(u_0-u)_{n'}^{s'} t^*_{\ve'_2}(u_0-v)_{s'}^n.
$$

\nsec Vertex Operator Approach

In the vertex operator approach (see Ref.~\Refs{\JMbook}
and references therein) the computation of correlation functions
is reduced to the algebraic problem of finding the traces of some
operators. Namely, the lattice is partitioned into several parts as it
is shown in Fig.~5. Then the partition functions of the parts with
fixed variables at their boundaries are treated as matrix elements
of some operators (corner transfer matrices\refs{\Baxter} and
vertex operators\refs{\DFJMN}) acting in the spaces associated with
a half-line.

\topinsert
%
%
\line{\hss
\beginpicture
\setcoordinatesystem units <1cm,1cm> point at 0 0
\rl 1.0 7.75 , 1.0 7.25 \rl 1.0 6.70 , 1.0 6.25
	\rl 1.0 5.20 , 1.0 4.75 \rl 1.0 4.25 , 1.0 3.6 \ar 1.0 3.61 , 1.0 3.6
\rl 1.5 8.25 , 1.5 7.25 \rl 1.5 6.70 , 1.5 6.25
	\rl 1.5 5.20 , 1.5 4.75 \rl 1.5 4.25 , 1.5 3.1 \ar 1.5 3.11 , 1.5 3.1
\rl 2.0 8.75 , 2.0 7.25 \rl 2.0 6.75 , 2.0 6.25
	\rl 2.0 5.25 , 2.0 4.75 \rl 2.0 4.25 , 2.0 2.6 \ar 2.0 2.61 , 2.0 2.6
\rl 2.5 9.25 , 2.5 7.25 \rl 2.5 6.75 , 2.5 6.25
	\rl 2.5 5.25 , 2.5 4.75 \rl 2.5 4.25 , 2.5 2.1 \ar 2.5 2.11 , 2.5 2.1
\rl 3.0 9.75 , 3.0 7.25 \rl 3.0 6.75 , 3.0 6.25
	\rl 3.0 5.25 , 3.0 4.75 \rl 3.0 4.25 , 3.0 1.6 \ar 3.0 1.61 , 3.0 1.6
\rl 4.0 9.75 , 4.0 7.25 \rl 4.0 6.75 , 4.0 6.25
	\rl 4.0 5.25 , 4.0 4.75 \rl 4.0 4.25 , 4.0 1.6 \ar 4.0 1.61 , 4.0 1.6
\rl 4.5 9.25 , 4.5 7.25 \rl 4.5 6.75 , 4.5 6.25
	\rl 4.5 5.25 , 4.5 4.75 \rl 4.5 4.25 , 4.5 2.1 \ar 4.5 2.11 , 4.5 2.1
\rl 5.0 8.75 , 5.0 7.25 \rl 5.0 6.70 , 5.0 6.25
	\rl 5.0 5.25 , 5.0 4.75 \rl 5.0 4.25 , 5.0 2.6 \ar 5.0 2.61 , 5.0 2.6
\rl 5.5 8.25 , 5.5 7.25 \rl 5.5 6.70 , 5.5 6.25
	\rl 5.5 5.20 , 5.5 4.75 \rl 5.5 4.25 , 5.5 3.1 \ar 5.5 3.11 , 5.5 3.1
\rl 6.0 7.75 , 6.0 7.25 \rl 6.0 6.70 , 6.0 6.25
	\rl 6.0 5.20 , 6.0 4.75 \rl 6.0 4.25 , 6.0 3.6 \ar 6.0 3.61 , 6.0 3.6
\setdots <2mm>
\rl 2.0 6.05 , 2.0 5.25
\rl 3.5 6.05 , 3.5 5.25
\rl 5.0 6.05 , 5.0 5.25
\setsolid
\rl 4.25 9.5 , 3.75 9.5 \rl 3.25 9.5 , 2.6 9.5 \ar 2.61 9.5 , 2.6 9.5
\rl 4.75 9.0 , 3.75 9.0 \rl 3.25 9.0 , 2.1 9.0 \ar 2.11 9.0 , 2.1 9.0
\rl 5.25 8.5 , 3.75 8.5 \rl 3.25 8.5 , 1.6 8.5 \ar 1.61 8.5 , 1.6 8.5
\rl 5.75 8.0 , 3.75 8.0 \rl 3.25 8.0 , 1.1 8.0 \ar 1.11 8.0 , 1.1 8.0
\rl 6.25 7.5 , 3.75 7.5 \rl 3.25 7.5 , 0.6 7.5 \ar 0.61 7.5 , 0.6 7.5
\rl 6.25 6.5 , 3.75 6.5 \rl 3.25 6.5 , 0.6 6.5 \ar 0.61 6.5 , 0.6 6.5
\rl 6.25 5.0 , 3.75 5.0 \rl 3.25 5.0 , 0.6 5.0 \ar 0.61 5.0 , 0.6 5.0
\rl 6.25 4.0 , 3.75 4.0 \rl 3.25 4.0 , 0.6 4.0 \ar 0.61 4.0 , 0.6 4.0
\rl 5.75 3.5 , 3.75 3.5 \rl 3.25 3.5 , 1.1 3.5 \ar 1.11 3.5 , 1.1 3.5
\rl 5.25 3.0 , 3.75 3.0 \rl 3.25 3.0 , 1.6 3.0 \ar 1.61 3.0 , 1.6 3.0
\rl 4.75 2.5 , 3.75 2.5 \rl 3.25 2.5 , 2.1 2.5 \ar 2.11 2.5 , 2.1 2.5
\rl 4.25 2.0 , 3.75 2.0 \rl 3.25 2.0 , 2.6 2.0 \ar 2.61 2.0 , 2.6 2.0
\put{$\ve_1$} [B] at 3.5 4.95
\put{$\ve_N$} [B] at 3.5 6.45
\put{$v_1$} [rB] at 0.5 4.95
\put{$v_N$} [rB] at 0.5 6.45
\put{$u$} [t] at 3.0 1.4
\put{$u$} [t] at 4.0 1.4
\put{$v$} [rB] at 0.5 3.95
\put{$v$} [rB] at 0.5 7.45
\put{$A(u-v)$} [lB] at 5.2 8.7
\put{$B(u-v)$} [rB] at 1.8 8.7
\put{$C(u-v)$} [rt] at 1.8 2.8
\put{$D(u-v)$} [lt] at 5.2 2.8
\put{$\Phi_{\ve_1}(u_1)$}   [rB] at 6.25 5.35
\put{$\Phi'_{\ve_1}(u_1)$} [lB] at 0.7  5.35
\put{$\Phi_{\ve_N}(u_N)$}   [rB] at 6.25 6.85
\put{$\Phi'_{\ve_N}(u_N)$} [lB] at 0.7  6.85
\put{$u_i=u-v_i$} [lB] at 5.5 1.3
\put{$(a)$} [B] at 3.5 0.0
\setcoordinatesystem units <1cm,1cm> point at -8 -0.5
\rl 1.0 7.5 , 1.0 7.0 \rl 1.0 6.5 , 1.0 6.0
	\rl 1.0 4.5 , 1.0 4.0 \rl 1.0 3.5 , 1.0 3.0
\rl 1.5 8.0 , 1.5 7.0 \rl 1.5 6.5 , 1.5 6.0
	\rl 1.5 4.5 , 1.5 4.0 \rl 1.5 3.5 , 1.5 2.5
\rl 2.0 8.5 , 2.0 7.0 \rl 2.0 6.5 , 2.0 6.0
	\rl 2.0 4.5 , 2.0 4.0 \rl 2.0 3.5 , 2.0 2.0
\rl 2.5 9.0 , 2.5 7.0 \rl 2.5 6.5 , 2.5 6.0
	\rl 2.5 4.5 , 2.5 4.0 \rl 2.5 3.5 , 2.5 1.5
\rl 3.0 9.0 , 3.0 7.0 \rl 3.0 6.5 , 3.0 6.0
	\rl 3.0 4.5 , 3.0 4.0 \rl 3.0 3.5 , 3.0 1.5
\rl 3.5 9.0 , 3.5 7.0 \rl 3.5 6.5 , 3.5 6.0
	\rl 3.5 4.5 , 3.5 4.0 \rl 3.5 3.5 , 3.5 1.5
\rl 4.0 9.0 , 4.0 7.0 \rl 4.0 6.5 , 4.0 6.0
	\rl 4.0 4.5 , 4.0 4.0 \rl 4.0 3.5 , 4.0 1.5
\rl 4.5 8.5 , 4.5 7.0 \rl 4.5 6.5 , 4.5 6.0
	\rl 4.5 4.5 , 4.5 4.0 \rl 4.5 3.5 , 4.5 2.0
\rl 5.0 8.0 , 5.0 7.0 \rl 5.0 6.5 , 5.0 6.0
	\rl 5.0 4.5 , 5.0 4.0 \rl 5.0 3.5 , 5.0 2.5
\rl 5.5 7.5 , 5.5 7.0 \rl 5.5 6.5 , 5.5 6.0
	\rl 5.5 4.5 , 5.5 4.0 \rl 5.5 3.5 , 5.5 3.0
\setdashes <2pt>
\rl 2.75 9.25 , 2.75 1.1
\rl 3.75 9.25 , 3.75 1.1
\setsolid
\ar 2.75 1.11 , 2.75 1.1
\ar 3.75 1.11 , 3.75 1.1
\setdots <2mm>
\rl 2.0  5.85 , 2.0  5.03
\rl 3.25 5.65 , 3.25 5.03
\rl 4.5  5.85 , 4.5  5.03
\setsolid
\rl 4.0 9.0 , 3.5 9.0 \rl 3.0 9.0 , 2.5 9.0
\rl 4.5 8.5 , 3.5 8.5 \rl 3.0 8.5 , 2.0 8.5
\rl 5.0 8.0 , 3.5 8.0 \rl 3.0 8.0 , 1.5 8.0
\rl 5.5 7.5 , 3.5 7.5 \rl 3.0 7.5 , 1.0 7.5
\rl 5.5 7.0 , 3.5 7.0 \rl 3.0 7.0 , 1.0 7.0
\rl 5.5 6.5 , 3.5 6.5 \rl 3.0 6.5 , 1.0 6.5
\rl 5.5 6.0 , 3.5 6.0 \rl 3.0 6.0 , 1.0 6.0
\rl 5.5 4.5 , 3.5 4.5 \rl 3.0 4.5 , 1.0 4.5
\rl 5.5 4.0 , 3.5 4.0 \rl 3.0 4.0 , 1.0 4.0
\rl 5.5 3.5 , 3.5 3.5 \rl 3.0 3.5 , 1.0 3.5
\rl 5.5 3.0 , 3.5 3.0 \rl 3.0 3.0 , 1.0 3.0
\rl 5.0 2.5 , 3.5 2.5 \rl 3.0 2.5 , 1.5 2.5
\rl 4.5 2.0 , 3.5 2.0 \rl 3.0 2.0 , 2.0 2.0
\rl 4.0 1.5 , 3.5 1.5 \rl 3.0 1.5 , 2.5 1.5
\setdashes <2pt>
\rl 5.75 7.25 , 0.6 7.25
\rl 5.75 6.25 , 0.6 6.25
\rl 5.75 4.25 , 0.6 4.25
\rl 5.75 3.25 , 0.6 3.25
\setsolid
\ar 0.61 7.25 , 0.6 7.25
\ar 0.61 6.25 , 0.6 6.25
\ar 0.61 4.25 , 0.6 4.25
\ar 0.61 3.25 , 0.6 3.25
\put{$n_0$}     [lB] at 3.1 3.95
\put{$n_1$}     [lB] at 3.1 4.45
\put{$n_{N-1}$} [lB] at 2.9 5.8
\put{$n_N$}     [lB] at 3.05 6.45
\put{$v_1$} [rB] at 0.5 4.2
\put{$v_2$} [rB] at 0.5 6.2
\put{$u$} [t] at 2.75 1.0
\put{$u$} [t] at 3.75 1.0
\put{$v$} [rB] at 0.5 3.2
\put{$v$} [rB] at 0.5 7.2
\put{$A(u-v)$} [lB] at 4.7 8.2
\put{$B(u-v)$} [rB] at 1.8 8.2
\put{$C(u-v)$} [rt] at 1.8 2.3
\put{$D(u-v)$} [lt] at 4.7 2.3
\put{$\Phi(u_1)^{n_1}_{n_0}$}   [rb] at 5.75 4.55
\put{$\Phi'(u_1)^{n_0}_{n_1}$} [lb] at 0.75 4.55
\put{$\Phi(u_N)^{n_N}_{n_{N-1}}$}   [rb] at 5.9  6.5
\put{$\Phi'(u_N)^{n_{N-1}}_{n_N}$} [lb] at 0.75 6.55
\put{$(b)$} [B] at 3.25 -0.5
\endpicture
\hss}
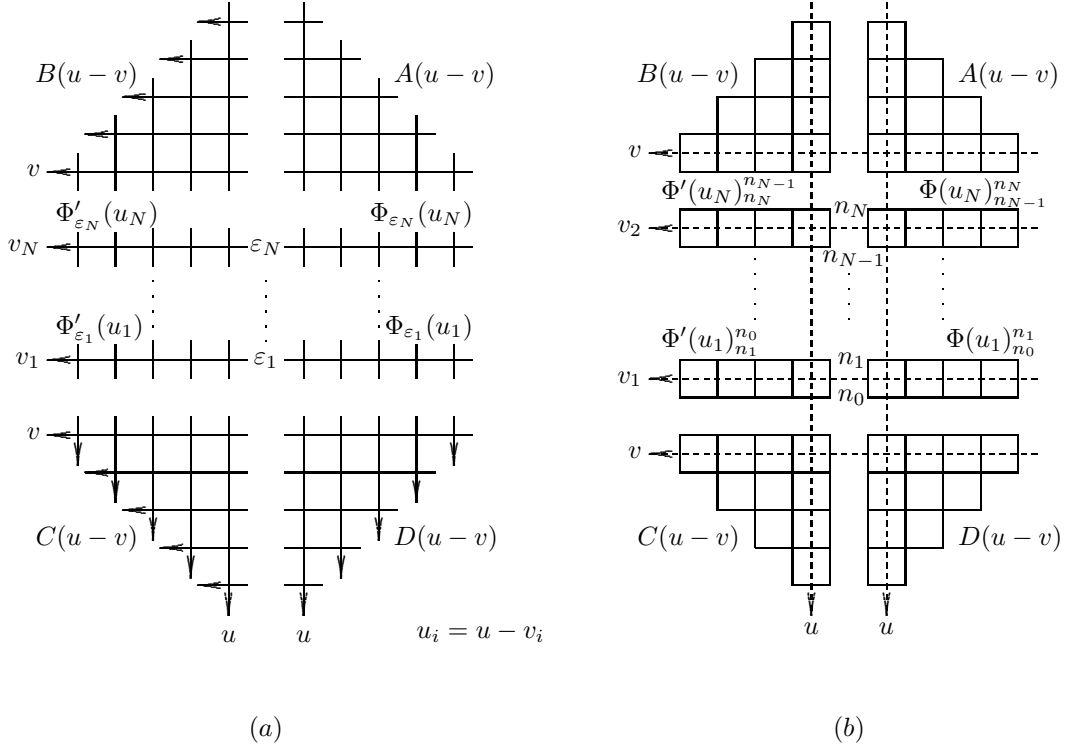
\figcap{5} Partition of the lattice into the corner transfer
matrices $A$, $B$, $C$, $D$ and vertex operators $\Phi$. The lattice
is only cut to stress the mode of partition, it must be considered
as an integral thing. Action of the operators is
anti-clockwise. $(a)$ Eight-vertex model. $(b)$ SOS model.\par
\endinsert

\ssubsec{1} Eight-Vertex Model

Let us begin with the eight-vertex model in the antiferroelectric
region\refs{\JMeightvertex}. Consider the subspace $\H_i$
in $\C^2\otimes\C^2\otimes\cdots$, spanned on the
vectors $|\ve_1\rangle\otimes|\ve_2\rangle\otimes\cdots$ such that
$\ve_k$ stabilizes to $(-)^{k+i}$ for large $k$. The operators
associated to the parts of the lattice, shown in Fig.~5a, are
$$
\eqalign{
\span
A^{(1-i,i)}(u),C^{(1-i,i)}(u):\H_i\rightarrow\H_{1-i},
\qquad
B^{(i)}(u),D^{(i)}(u):\H_i\rightarrow\H_i,
\cr\span
\Phi_\ve^{(1-i,i)}(u),\Phi_\ve^{(1-i,i)\prime}(u):\H_i\rightarrow\H_{1-i}.
}
$$

The product of
four corner transfer matrices in the infinite lattice is
an operator independent of $u$:
$$
\rho^{(i)}\equiv
D^{(i)}(u)C^{(i,1-i)}(u)B^{(1-i)}(u)A^{(1-i,i)}(u)
=\const x^{4H^{(i)}},
\eqlabel\EQevrho
$$
where $\const$ is a $c$-number and $H^{(i)}$
is so called corner Hamiltonian.

The spectrum of the operator $H^{(i)}$ is equidistant.
More precisely, it is determined by the generating
function
$$
\chi^{(i)}(q)\equiv\Tr_{\H_i}q^{H^{(i)}}=(q^{1/2};q)_\infty^{-1}.
\eqlabel\EQevcharacter
$$

The vertex operators $\Phi^{(1-i,i)}_\ve(u)$ satisfy the following
relations which are consequences of the Yang--Baxter equation,
unitarity and crossing symmetry:
$$
\eqalignno{
\span
\Phi^{(i,1-i)}_{\ve_1}(u_1)\Phi^{(1-i,i)}_{\ve_2}(u_2)
=\sum_{\ve'_1\ve'_2}R(u_1-u_2)^{\ve'_1\ve'_2}_{\ve_1\ve_2}
\Phi^{(i,1-i)}_{\ve'_2}(u_2)\Phi^{(1-i,i)}_{\ve'_1}(u_1),
\lnlabel\EQevcommutation
\cr
\span
\Phi^{(1-i,i)}_\ve(u)\rho^{(i)}=\rho^{(1-i)}\Phi^{(1-i,i)}_\ve(u-2),
\lnlabel\EQevdiff
\cr
\span
\sum_\ve\Phi^{(i,1-i)}_{-\ve}(u-1)\Phi^{(1-i,i)}_\ve(u)=1.
\lnlabel\EQevnormalization}
$$
The conjugate vertex operator $\Phi^{(i,1-i)*}_\ve(u)$
defined as
$$
\Phi^{(1-i,i)\prime}_\ve(u)B^{(i)}(u)A^{(i,1-i)}(u)
=B^{(1-i)}(u)A^{(1-i,i)}(u)\Phi^{(i,1-i)*}_\ve(u),
\qquad
\sum_\ve\Phi^{(i,1-i)*}_\ve(u)\Phi^{(1-i,i)}_\ve(u)=1
\eqlabel\EQevconjugatedef
$$
is given by
$$
\Phi^{(i,1-i)*}_\ve(u)=\Phi^{(i,1-i)}_{-\ve}(u-1).
\eqlabel\EQevconjugate
$$
The probability of the configuration of polarizations on an inhomogeneous
lattice as it is shown in Fig.~5a is given by
$$
\eqalign{
\span
P^{(i)}_{\ve_1\ldots\ve_N}
={1\over\chi^{(i)}(x^4)}
\Tr_{\H_i}\left(
\Phi^{(i,1-i)*}_{\ve_1}(u_1)\ldots\Phi^{(1-i',i')*}_{\ve_N}(u_N)
\Phi^{(i',1-i')}_{\ve_N}(u_N)\ldots\Phi^{(1-i,i)}_{\ve_1}(u_1)\,
x^{4H^{(i)}}\right),
\cr
\span
u_k=u-v_k,\qquad (-)^{i'}=(-)^{N+i}.
}\eqlabel\EQevcorrfun
$$

Consider more general functions
$$
F^{(i)}_{\ve_1\ldots\ve_N}(u_1,\ldots,u_N)
={1\over\chi^{(i)}(x^4)}\Tr_{\H_i}\left(
\Phi^{(i,1-i)}_{\ve_N}(u_N)\ldots\Phi^{(1-i,i)}_{\ve_1}(u_1)\,
x^{4H^{(i)}}\right)
\eqlabel\EQevtrace
$$
for even $N$.
They satisfy the equations\refs{\JMeightvertex}
$$
\eqadvance\EQevkz
\eqalignno{
\span
F^{(i)}_{\ve_1\ldots\ve_N}(\ldots,u_j+{2\pi\i\over\epsilon},\ldots)
=F^{(i)}_{\ve_1\ldots\ve_N}(\ldots,u_j,\ldots),
\lnlabelno(\EQevkz a)
\cr
\span
F^{(i)}_{\ve_1\ldots\ve_N}(u_1+a,\ldots,u_N+a)
=F^{(i)}_{\ve_1\ldots\ve_N}(u_1,\ldots,u_N)
\lnlabelno(\EQevkz b)
\cr
\smash{\sum_{\ve'_j\ve'_{j+1}}}
R(u_j-u_{j+1})^{\ve'_j\ve'_{j+1}}_{\ve_j\ve_{j+1}}
F^{(i)}_{\ldots\ve'_j\ve'_{j+1}\ldots}(\ldots,u_j,u_{j+1},\ldots)&
\cr
&\hskip -5em
=F^{(i)}_{\ldots\ve_{j+1}\ve_j\ldots}(\ldots,u_{j+1},u_j,\ldots),
\lnlabelno(\EQevkz c)
\cr
\span
F^{(i)}_{\ve_1\ve_2\ldots\ve_N}(u_1+2,u_2,\ldots,u_N)
=F^{(1-i)}_{\ve_2\ldots\ve_N\ve_1}(u_2,\ldots,u_N,u_1),
\lnlabelno(\EQevkz d)
\cr
\span
\sum_\ve F^{(i)}_{\ve_1\ldots\ve_N,\ve,-\ve}(u_1,\ldots,u_N,u,u-1)
=F^{(i)}_{\ve_1\ldots\ve_N}(u_1,\ldots,u_N),
\qquad
\sum_\ve F^{(i)}_{-\ve,\ve}(u,u-1)=1.
\lnlabelno(\EQevkz e)}
$$
These equations follow from Eqs.~(\EQevdiff--\EQevnormalization) and
can be considered as defining equations for correlation functions.

\ssubsec{2} SOS model

For the SOS model the construction repeats that
for the eight-vertex model in general outline (see Fig.~5b).
The corner transfer matrices are
labelled by two integers $m$ and $n$. The variable $n$ denotes the
height at the corner. The number $m$ describes the condition
at the infinity. Namely, for the
space $\H_{m,n}$ with $0<m<r-1$ is the subspace of
$\C^\infty\otimes\C^\infty\otimes\cdots$
spanned of vectors
$|n_0\rangle\otimes|n_1\rangle\otimes|n_2\rangle\otimes\cdots$,
such that $n_0=n$, each pair $(n_k,n_{k+1})$ is an admissible pair of heights,
and for $k$ large enough $n_{2k}=m$ and $n_{2k+1}=m+1$ if $n-m$ is even,
$n_{2k}=m+1$ and $n_{2k+1}=m$ if $n-m$ is odd.
The situation of general $m$ is more complicated, but we need not
discuss it for our purposes. For simplicity we restrict the discussion
to $r>2$, but our final construction is valid for any $r>1$.
Parts of the lattice shown in Fig.~5b define again operators:
$$
\eqalign{
\span
A_{m,n}(u),B_{m,n}(u),C_{m,n}(u),D_{m,n}(u):\H_{m,n}\rightarrow\H_{m,n},
\cr\span
\Phi(u)^{m,n'}_{m,n},\Phi'(u)^{m,n'}_{m,n}:\H_{m,n}\rightarrow\H_{m,n'}.
}
$$
The product of the corner transfer matrices is again a
$u$-independent operator:
$$
\rho_{m,n}\equiv D_{m,n}(u)C_{m,n}(u)B_{m,n}(u)A_{m,n}(u)
=\const[n]\,x^{4H_{m,n}}.
\eqlabel\EQsosrho
$$
The operator $H_{m,n}$ also possesses an equidistant spectrum.
We assume that the respective generating function is
$$
\chi_{m,n}(q)\equiv\Tr_{\H_{m,n}}q^{H_{m,n}}
=q^{(rm-(r-1)n)^2/4r(r-1)}(q;q)_\infty^{-1}.
\eqlabel\EQsoscharacter
$$
The vertex operators $\Phi(u)^{n'}_n$ (we usually omit indices $m$
in operators from here on) obey the relations:
$$
\eqalignno{
\span
\Phi(u_1)^{n'}_s\Phi(u_2)^s_n
=\sum_{s'}\W(s,n;s',n'|u_1-u_2)
\Phi(u_2)^{n'}_{s'}\Phi(u_1)^{s'}_n,
\lnlabel\EQsoscommutation
\cr
\span
\Phi(u)^{n'}_n\rho_n=\rho_{n'}\Phi(u-2)^{n'}_n,
\lnlabel\EQsosdiff
\cr
\span
\sum_{n'}(n'-n)[n']\,\Phi(u-1)^n_{n'}\Phi(u)^{n'}_n=(-1)^{n-m}.
\lnlabel\EQsosnormalization}
$$
The conjugate operator defined as
$$
\Phi'(u)^{n'}_n B_n(u)A_n(u)
=B_{n'}(u)A_{n'}(u)\Phi^*(u)^{n'}_n,
\qquad
\sum_{n'}\Phi^*(u)^n_{n'}\Phi(u)^{n'}_n=1
\eqlabel\EQsosconjugatedef
$$
is given by
$$
\Phi^*(u)^{n'}_n=(-)^{n-m}(n'-n)[n]\,\Phi(u-1)^{n'}_n.
\eqlabel\EQsosconjugate
$$

\nsec Vertex-Face Correspondence and Correlation Functions

Now we are ready to explore the vertex-face correspondence for
the vertex operator construction.

Let us enclose a piece of lattice of the eight-vertex model
with several fixed polarizations with the intertwining vectors from
the upper and right sides and with the conjugate intertwining vectors
from the lower and left sides (Fig.~6a). We can shrink such loop using
the vertex-face correspondence and the definition of the conjugate
intertwining vector (\EQstarprimedef) (Fig.~6b). But the tail
in the right lower part of Fig.~6b cannot be cancelled, because
the heights $n$ and $n'$ designated at the figure are generically
not equal.

\topinsert
%
%
\line{\hss
\beginpicture
\setcoordinatesystem units <1cm,1cm> point at 0 0
\trl 5.5 7.5 , 5.5 11.5 \trl 5.5 11.5 , 1.5 11.5
\trl 1.5 11.5 , 1.5 7.5 \trl 1.5 7.5 , 5.5 7.5
\ln 5.5 7.9 , 5.4 8.0 \ln 5.4 8.0 , 5.5 8.1
\linethickness=.8pt
\rl 5.48 7.92 , 5.48 8.08 \rl 5.46 7.94 , 5.46 8.06
\rl 5.44 7.96 , 5.44 8.04 \rl 5.42 7.98 , 5.42 8.02
\linethickness=.4pt
\ln 5.5 8.9 , 5.4 9.0 \ln 5.4 9.0 , 5.5 9.1
\ln 5.5 9.9 , 5.4 10.0 \ln 5.4 10.0 , 5.5 10.1
\ln 5.5 10.9 , 5.4 11.0 \ln 5.4 11.0 , 5.5 11.1
\ln 5.1 11.5 , 5.0 11.4 \ln 5.0 11.4 , 4.9 11.5
\ln 4.1 11.5 , 4.0 11.4 \ln 4.0 11.4 , 3.9 11.5
\ln 3.1 11.5 , 3.0 11.4 \ln 3.0 11.4 , 2.9 11.5
\ln 2.1 11.5 , 2.0 11.4 \ln 2.0 11.4 , 1.9 11.5
\ln 1.5 11.1 , 1.6 11.0 \ln 1.6 11.0 , 1.5 10.9
\ln 1.5 10.1 , 1.6 10.0 \ln 1.6 10.0 , 1.5 9.9
\ln 1.5 9.1 , 1.6 9.0 \ln 1.6 9.0 , 1.5 8.9
\ln 1.5 8.1 , 1.6 8.0 \ln 1.6 8.0 , 1.5 7.9
\ln 1.9 7.5 , 2.0 7.6 \ln 2.0 7.6 , 2.1 7.5
\ln 2.9 7.5 , 3.0 7.6 \ln 3.0 7.6 , 3.1 7.5
\ln 3.9 7.5 , 4.0 7.6 \ln 4.0 7.6 , 4.1 7.5
\ln 4.9 7.5 , 5.0 7.6 \ln 5.0 7.6 , 5.1 7.5
\rl 2.0 11.4 , 2.0 7.6
\rl 3.0 11.4 , 3.0 7.6
\rl 4.0 11.4 , 4.0 7.6
\rl 5.0 11.4 , 5.0 7.6
\rl 5.4 11.0 , 1.6 11.0
\rl 5.4 10.0 , 1.6 10.0
\rl 5.4 9.0 , 1.6 9.0
\rl 5.4 8.0 , 1.6 8.0
\setdashes <2.5pt>
\rl 2.0 7.5 , 2.0 6.9
\rl 3.0 7.5 , 3.0 6.9
\rl 4.0 7.5 , 4.0 6.9
\rl 5.0 7.5 , 5.0 6.9
\rl 1.5 11.0 , 0.9 11.0
\rl 1.5 10.0 , 0.9 10.0
\rl 1.5 9.0 , 0.9 9.0
\rl 1.5 8.0 , 0.9 8.0
\setsolid
\ar 2.0 6.91 , 2.0 6.9
\ar 3.0 6.91 , 3.0 6.9
\ar 4.0 6.91 , 4.0 6.9
\ar 5.0 6.91 , 5.0 6.9
\ar 0.91 11.0 , 0.9 11.0
\ar 0.91 10.0 , 0.9 10.0
\ar 0.91 9.0 , 0.9 9.0
\ar 0.91 8.0 , 0.9 8.0
\bfpoint 3.5 9.0
\bfpoint 3.5 10.0
\put{$\ve_1$} [lB] at 3.5 9.1
\put{$\ve_2$} [lB] at 3.5 10.1
\put{$(a)$} [B] at 3.5 6.5
%
\rl 1.0 6.0 , 1.0 3.0 \rl 1.0 2.0 , 1.0 1.0
\rl 2.0 6.0 , 2.0 3.0 \rl 2.0 2.0 , 2.0 1.0
\rl 3.0 6.0 , 3.0 5.0 \trl 3.0 5.0 , 3.0 3.0 \rl 3.0 2.0 , 3.0 1.0
\rl 4.0 6.0 , 4.0 5.0 \trl 4.0 5.0 , 4.0 3.0 \rl 4.0 2.0 , 4.0 1.0
\rl 5.0 6.0 , 5.0 3.0 \rl 5.0 2.0 , 5.0 1.0
\rl 6.0 6.0 , 6.0 3.0 \rl 6.0 2.0 , 6.0 1.0
\rl 6.0 6.0 , 4.0 6.0 \rl 3.0 6.0 , 1.0 6.0
\rl 6.0 5.0 , 4.0 5.0 \rl 3.0 5.0 , 1.0 5.0
\rl 6.0 4.0 , 4.0 4.0 \rl 3.0 4.0 , 1.0 4.0
\trl 6.0 3.0 , 4.0 3.0 \rl 3.0 3.0 , 1.0 3.0
\trl 6.0 2.0 , 4.0 2.0 \rl 3.0 2.0 , 1.0 2.0
\rl 6.0 1.0 , 4.0 1.0 \rl 3.0 1.0 , 1.0 1.0
\ln 4.0 4.6 , 3.9 4.5 \ln 3.9 4.5 , 4.0 4.4
\ln 3.0 4.6 , 3.1 4.5 \ln 3.1 4.5 , 3.0 4.4
\rl 3.9 4.5 , 3.1 4.5
\bfpoint 3.5 4.5
\ln 4.0 3.6 , 3.9 3.5 \ln 3.9 3.5 , 4.0 3.4
\ln 3.0 3.6 , 3.1 3.5 \ln 3.1 3.5 , 3.0 3.4
\rl 3.9 3.5 , 3.1 3.5
\bfpoint 3.5 3.5
\ln 4.4 3.0 , 4.5 2.9 \ln 4.5 2.9 , 4.6 3.0
\ln 4.4 2.0 , 4.5 2.1 \ln 4.5 2.1 , 4.6 2.0
\rl 4.5 2.9 , 4.5 2.1
\ln 5.4 3.0 , 5.5 2.9 \ln 5.5 2.9 , 5.6 3.0
\ln 5.4 2.0 , 5.5 2.1 \ln 5.5 2.1 , 5.6 2.0
\rl 5.5 2.9 , 5.5 2.1
\setdots <2.5pt>
\rl 4.0 6.0 , 3.0 6.0
\rl 4.0 5.0 , 3.0 5.0
\rl 4.0 2.0 , 3.0 2.0
\rl 4.0 1.0 , 3.0 1.0
\rl 1.0 3.0 , 1.0 2.0
\rl 2.0 3.0 , 2.0 2.0
\rl 3.0 3.0 , 3.0 2.0
\setdashes <2.5pt>
\rl 5.5 6.3 , 5.5 3.0 \rl 5.5 2.0 , 5.5 0.4
\rl 4.5 6.3 , 4.5 3.0 \rl 4.5 2.0 , 4.5 0.4
\rl 2.5 6.3 , 2.5 0.4
\rl 1.5 6.3 , 1.5 0.4
\rl 6.3 5.5 , 0.4 5.5
\rl 6.3 4.5 , 4.0 4.5 \rl 3.0 4.5 , 0.4 4.5
\rl 6.3 3.5 , 4.0 3.5 \rl 3.0 3.5 , 0.4 3.5
\rl 6.3 1.5 , 0.4 1.5
\setsolid
\ar 5.5 0.41 , 5.5 0.4
\ar 4.5 0.41 , 4.5 0.4
\ar 2.5 0.41 , 2.5 0.4
\ar 1.5 0.41 , 1.5 0.4
\ar 0.41 5.5 , 0.4 5.5
\ar 0.41 4.5 , 0.4 4.5
\ar 0.41 3.5 , 0.4 3.5
\ar 0.41 1.5 , 0.4 1.5
\put{$\ve_1$} [lB] at 3.5 3.7
\put{$\ve_2$} [lB] at 3.5 4.7
\put{$n$}  [rB] at 3.9 2.1
\put{$n\smash{{}'}$} [rt] at 3.9 2.9
\put{$(b)$} [B] at 3.5 0.0
\setcoordinatesystem units <1cm,1cm> point at -7 -0.5
\rl 1.0 8.5 , 1.0 8.0 \rl 1.0 7.5 , 1.0 7.0
	\rl 1.0 5.5 , 1.0 5.0 \rl 1.0 3.5 , 1.0 3.0
\rl 1.5 9.0 , 1.5 8.0 \rl 1.5 7.5 , 1.5 7.0
	\rl 1.5 5.5 , 1.5 5.0 \rl 1.5 3.5 , 1.5 2.5
\rl 2.0 9.5 , 2.0 8.0 \rl 2.0 7.5 , 2.0 7.0
	\rl 2.0 5.5 , 2.0 5.0 \rl 2.0 3.5 , 2.0 2.0
\rl 2.5 10.0 , 2.5 8.0 \rl 2.5 7.5 , 2.5 7.0
	\rl 2.5 5.5 , 2.5 5.0 \rl 2.5 3.5 , 2.5 1.5
\rl 3.0 10.0 , 3.0 8.0 \rl 3.0 3.5 , 3.0 1.5
\rl 4.0 10.0 , 4.0 8.0 \rl 4.0 3.5 , 4.0 1.5
\rl 4.5 10.0 , 4.5 8.0 \rl 4.5 7.5 , 4.5 7.0
	\rl 4.5 5.5 , 4.5 5.0 \rl 4.5 3.5 , 4.5 1.5
\rl 5.0 9.5 , 5.0 8.0 \rl 5.0 7.5 , 5.0 7.0
	\rl 5.0 5.5 , 5.0 5.0 \rl 5.0 3.5 , 5.0 2.0
\rl 5.5 9.0 , 5.5 8.0 \rl 5.5 7.5 , 5.5 7.0
	\rl 5.5 5.5 , 5.5 5.0 \rl 5.5 3.5 , 5.5 2.5
\rl 6.0 8.5 , 6.0 8.0 \rl 6.0 7.5 , 6.0 7.0
	\rl 6.0 5.5 , 6.0 5.0 \rl 6.0 3.5 , 6.0 3.0
\rl 2.5 10.0 , 3.0 10.0 \rl 4.0 10.0 , 4.5 10.0
\rl 2.0 9.5 , 3.0 9.5 \rl 4.0 9.5 , 5.0 9.5
\rl 1.5 9.0 , 3.0 9.0 \rl 4.0 9.0 , 5.5 9.0
\rl 1.0 8.5 , 3.0 8.5 \rl 4.0 8.5 , 6.0 8.5
\rl 1.0 8.0 , 3.0 8.0 \rl 4.0 8.0 , 6.0 8.0
\rl 1.0 7.5 , 3.0 7.5 \rl 4.0 7.5 , 6.0 7.5
\rl 1.0 7.0 , 3.0 7.0 \rl 4.0 7.0 , 6.0 7.0
\rl 1.0 5.5 , 3.0 5.5 \rl 4.0 5.5 , 6.0 5.5
\rl 1.0 5.0 , 3.0 5.0 \rl 4.0 5.0 , 6.0 5.0
\rl 1.0 3.5 , 3.0 3.5 \rl 4.0 3.5 , 6.0 3.5
\rl 1.0 3.0 , 3.0 3.0 \rl 4.0 3.0 , 6.0 3.0
\rl 1.5 2.5 , 3.0 2.5 \rl 4.0 2.5 , 5.5 2.5
\rl 2.0 2.0 , 3.0 2.0 \rl 4.0 2.0 , 5.0 2.0
\rl 2.5 1.5 , 3.0 1.5 \rl 4.0 1.5 , 4.5 1.5
\trl 4.0 7.5 , 4.0 7.0
\ln 4.0 7.15 , 3.9 7.25 \ln 3.9 7.25 , 4.0 7.35
\rl 3.9 7.25 , 3.75 7.25
\trl 3.0 7.5 , 3.0 7.0
\ln 3.0 7.15 , 3.1 7.25 \ln 3.1 7.25 , 3.0 7.35
\rl 3.1 7.25 , 3.25 7.25
\trl 4.0 5.5 , 4.0 5.0
\ln 4.0 5.15 , 3.9 5.25 \ln 3.9 5.25 , 4.0 5.35
\rl 3.9 5.25 , 3.7 5.25
\trl 3.0 5.5 , 3.0 5.0
\ln 3.0 5.15 , 3.1 5.25 \ln 3.1 5.25 , 3.0 5.35
\rl 3.1 5.25 , 3.3 5.25
\trl 4.0 4.5 , 6.0 4.5
\trl 4.0 4.0 , 6.0 4.0
\ln 4.15 4.5 , 4.25 4.4 \ln 4.25 4.4 , 4.35 4.5
\ln 4.65 4.5 , 4.75 4.4 \ln 4.75 4.4 , 4.85 4.5
\ln 5.15 4.5 , 5.25 4.4 \ln 5.25 4.4 , 5.35 4.5
\ln 5.65 4.5 , 5.75 4.4 \ln 5.75 4.4 , 5.85 4.5
\ln 4.15 4.0 , 4.25 4.1 \ln 4.25 4.1 , 4.35 4.0
\ln 4.65 4.0 , 4.75 4.1 \ln 4.75 4.1 , 4.85 4.0
\ln 5.15 4.0 , 5.25 4.1 \ln 5.25 4.1 , 5.35 4.0
\ln 5.65 4.0 , 5.75 4.1 \ln 5.75 4.1 , 5.85 4.0
\rl 4.25 4.4 , 4.25 4.1
\rl 4.75 4.4 , 4.75 4.1
\rl 5.25 4.4 , 5.25 4.1
\rl 5.75 4.4 , 5.75 4.1
\setdots <2pt>
\rl 3.0 8.0 , 3.0 7.5 \rl 3.0 7.0 , 3.0 6.7
	\rl 3.0 5.8 , 3.0 5.5 \rl 3.0 5.0 , 3.0 3.5
\rl 4.0 8.0 , 4.0 7.5 \rl 4.0 7.0 , 4.0 6.7
	\rl 4.0 5.8 , 4.0 5.5 \rl 4.0 5.0 , 4.0 4.5 \rl 4.0 4.0 , 4.0 3.5
\rl 3.0 8.0 , 4.0 8.0
\rl 3.0 3.5 , 4.0 3.5
\setdots <2mm>
\rl 2.0 6.7 , 2.0 5.9
\rl 3.5 6.7 , 3.5 5.9
\rl 5.0 6.7 , 5.0 5.9
\setdashes <2pt>
\rl 2.75 10.4 , 2.75 1.1
\rl 3.5 10.2 , 3.5 7.5 \rl 3.5 5.0 , 3.5 4.25 \rl 3.5 4.25 , 6.2 4.25
\rl 4.25 10.2 , 4.25 4.5 \rl 4.25 4.0 , 4.25 1.1
\rl 6.2 8.25 , 0.6 8.25
\rl 6.2 7.25 , 4.0 7.25 \rl 3.0 7.25 , 0.6 7.25
\rl 6.2 5.25 , 4.0 5.25 \rl 3.0 5.25 , 0.6 5.25
\rl 6.2 3.25 , 0.6 3.25
\setsolid
\ar 2.75 1.11 , 2.75 1.1
\ar 4.25 1.11 , 4.25 1.1
\ar 0.61 8.25 , 0.6 8.25
\ar 0.61 7.25 , 0.6 7.25
\ar 0.61 5.25 , 0.6 5.25
\ar 0.61 3.25 , 0.6 3.25
\ar 3.5 10.39 , 3.5 10.4
\put{$\ve_1$} [B] at 3.55 5.2
\put{$\ve_N$} [B] at 3.55 7.2
\put{$v_1$} [rB] at 0.5 5.2
\put{$v_N$} [rB] at 0.5 7.2
\put{$u$}  [t] at 4.25 1.0
\put{$u$}  [t] at 2.75 1.0
\put{$v$} [rB] at 0.5 3.2
\put{$v$} [rB] at 0.5 8.2
\put{$v_0$} [B] at 3.5 10.5
\put{$A(u-v)$} [lB] at 5.2 9.2
\put{$B(u-v)$} [rB] at 1.8 9.2
\put{$C(u-v)$} [rt] at 1.8 2.3
\put{$D(u-v)$} [lt] at 5.2 2.3
\put{$\Phi_{\ve_1}(u;u_0)$}   [rB] at 6.5 5.65
\put{$\Phi'_{\ve_1}(u;u_0)$} [lB] at 0.5 5.65
\put{$\Phi_{\ve_1}(u;u_0)$}   [rB] at 6.5 7.65
\put{$\Phi'_{\ve_1}(u;u_0)$} [lB] at 0.5 7.65
\put{$\Lambda(u_0)$} [rB] at 6.2 4.65
\put{$u_i=u-v_i$} [lB] at 5.0 1.0
\put{$u_0=u-v_0$} [lB] at 5.0 0.6
\put{$(c)$} [B] at 3.5 -0.5
\endpicture
\hss}
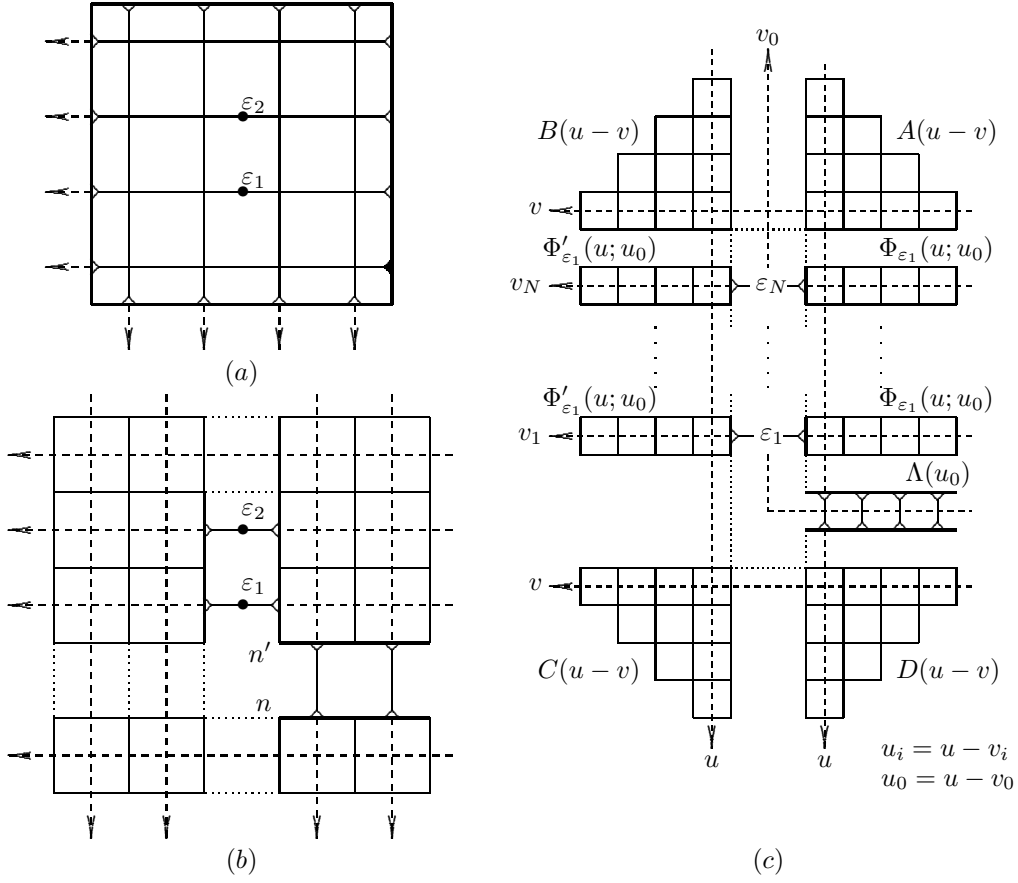
\figcap{6} Contraction of the loop of intertwining vectors
in the lattice with insertions. $(a)$ Initial loop. The $t'$ vector
in the lower write side is used to simplify the next figure.
$(b)$ `Maximal' contraction. Dotted lines connect points
with coincident heights. Summation in $n$ and $n'$ is implied,
so they do not necessarily coincide. $(c)$ Partition of the lattice
after such contraction.\par
\endinsert

Now let us establish correspondence between ground states (or
conditions at infinity) for the SOS and eight-vertex models.
Let $m$ in the definition of the intertwining vectors (\EQintertwining)
coincide with $m$ characterizing the condition at the infinity.
To do it, consider the `low temperature limit' $x\to0$
($\epsilon\to\infty$). For simplicity we consider
the case $0<m<r-1$.

It can be checked (see Appendix A) that
the relevant values of $u$ in $t_\ve(u)^{n'}_n$ and
$t^*_\ve(u)^{n'}_n$ for the free field representation are
$$
-2<\Re u<0.
$$
In the limit $x\to0$ we have for the intertwining
vectors
$$
\eqalign{
t_+(u)^{n'}_n
&\simeq\sqrt{\epsilon r\over\pi}\,\e^{-{\epsilon\over2r}((n'-n)u+n')^2},
\cr
t_-(u)^{n'}_n
&\simeq(-)^{n-m+1}
\sqrt{\epsilon r\over\pi}\,\e^{-{\epsilon\over2r}((n'-n)u+n')^2
-\epsilon r+\epsilon((n'-n)u+n')},
\cr
t^*_+(u)^{n'}_n
&\simeq(n-n')\sqrt{\pi\over\epsilon r}
\,\e^{{\epsilon\over2r}((n-n')u+n)^2+(1+n'-n)\epsilon u},
\cr
t^*_-(u)^{n'}_n
&\simeq(-)^{n-m+1}(n'-n)\sqrt{\pi\over\epsilon r}
\,\e^{{\epsilon\over2r}((n-n')u+n)^2+\epsilon r-\epsilon(n-u)},
\qquad n,n'>0.
}\eqlabel\EQlowtemperature
$$
Consider now the quantities (Fig.~7a)
$$
a^{(1-i,i)}_\ve
=t^*_\ve(u)^{m+i}_{m+1-i} t_\ve(u)^{m+1-i}_{m+i}.
\eqlabel\EQaquantities
$$
It is easy to check that
$$
|a^{(1,0)}_+|\simeq1,
\qquad
|a^{(1,0)}_-|\simeq\e^{2\epsilon\Re u},
\qquad
|a^{(0,1)}_+|\simeq\e^{2\epsilon\Re u},
\qquad
|a^{(0,1)}_-|\simeq1.
$$
We see that
$$
|a^{(1,0)}_+|\gg|a^{(1,0)}_-|,
\qquad
|a^{(0,1)}_-|\gg|a^{(0,1)}_+|.
$$
The configurations shown in Fig.~7b give the leading contribution
into the partition function. The other two configurations
are vanishing in the low-temperature limit. It means that
a definite ground state configuration of the SOS model imposes
a definite ground state configuration of the eight-vertex model.
Hence the SOS condition at the infinity $m$, such that
$n-m\in2\Z+i$, imposes the
condition at the infinity $i$ for the eight-vertex model.

\topinsert
%
%
\line{\hfil
\beginpicture
\setcoordinatesystem units <1cm,1cm> point at 0 0
\trl 2.5 2.5 , 4.0 2.5
\ln 3.15 2.5 , 3.25 2.4 \ln 3.25 2.4 , 3.35 2.5
\trl 2.5 1.0 , 4.0 1.0
\ln 3.15 1.0 , 3.25 1.1 \ln 3.25 1.1 , 3.35 1.0
\rl 3.25 2.4 , 3.25 1.1
\setdashes <2.5pt>
\rl 3.25 1.0 , 3.25 0.4
\setsolid
\ar 3.25 0.41 , 3.25 0.4
\put{$\ve$} [l] at 3.4 1.75
\put{$m+i$} [lB] at 3.7 2.7
\put{$m+i$} [lt] at 3.7 0.8
\put{$m+1-i$} [rB] at 3.0 2.7
\put{$m+1-i$} [rt] at 3.0 0.8
\put{\fourteenpoint$a\smash{{}^{(1-i,i)}_\ve}\ =$} [r] at 1.6 1.75
\put{$(a)$} [B] at 2.25 0.0
\setcoordinatesystem units <1cm,1cm> point at -4.5 0
\trl 2.5 2.5 , 4.0 2.5
\ln 3.15 2.5 , 3.25 2.4 \ln 3.25 2.4 , 3.35 2.5
\trl 2.5 1.0 , 4.0 1.0
\ln 3.15 1.0 , 3.25 1.1 \ln 3.25 1.1 , 3.35 1.0
\rl 3.25 2.4 , 3.25 1.1
\setdashes <2.5pt>
\rl 3.25 1.0 , 3.25 0.4
\setsolid
\ar 3.25 0.41 , 3.25 0.4
\put{$+$} [l] at 3.4 1.75
\put{$m$} [lB] at 3.9 2.7
\put{$m$} [lt] at 3.9 0.8
\put{$m+1$} [rB] at 2.8 2.7
\put{$m+1$} [rt] at 2.8 0.8
\setcoordinatesystem units <1cm,1cm> point at -7.5 0
\trl 2.5 2.5 , 4.0 2.5
\ln 3.15 2.5 , 3.25 2.4 \ln 3.25 2.4 , 3.35 2.5
\trl 2.5 1.0 , 4.0 1.0
\ln 3.15 1.0 , 3.25 1.1 \ln 3.25 1.1 , 3.35 1.0
\rl 3.25 2.4 , 3.25 1.1
\setdashes <2.5pt>
\rl 3.25 1.0 , 3.25 0.4
\setsolid
\ar 3.25 0.41 , 3.25 0.4
\put{$-$} [l] at 3.4 1.75
\put{$m+1$} [lB] at 3.7 2.7
\put{$m+1$} [lt] at 3.7 0.8
\put{$m$} [rB] at 2.6 2.7
\put{$m$} [rt] at 2.6 0.8
\put{$(b)$} [B] at 2.0 0.0
\endpicture
\hfil}
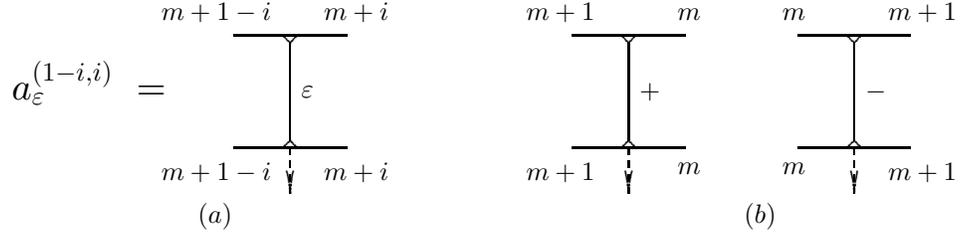
\figcap{7} Vertex-face correspondence of ground states:
$(a)$ basic weights; $(b)$ configurations of maximal weight
at low temperatures ($x\to0$) for $m>0$.\par
\endinsert

So we can partition the lattice as it is shown in Fig.~6c.
According to this partition we introduce the following objects:
the corner transfer matrices of the SOS model,
the vertex operators, and the `tail operator'.

Let us begin with the corner transfer matrices
of the SOS model
$$
\rho^{\prime(i)}
=\bigoplus_{n\in2\Z+m+i} D_{m,n}(u)C_{m,n}(u)B_{m,n}(u)A_{m,n}(u)
=\const\bigoplus_{n\in2\Z+m+i}[n]\,x^{4H_n}.
\eqlabel\EQvfcorner
$$
We assume that
one have to keep fixed an arbitrary
nonzero value of $m$ to calculate correlation functions.
This conjecture seems to be
physically reasonable, because such condition is compatible with
the vertex-face correspondence: we may take respective values of
heights at the boundary of Fig.~6a, not disturbing the shrinking
procedure. Besides, the correlation function in the eight-vertex model
must be sensitive to the eight-vertex boundary condition labeled by
$i\equiv n-m\pmod2$, but insensitive to details of arrangement
of the facilities that create these conditions, like a system of
intertwining vectors and the SOS lattice somewhere at infinity.

The vertex operators are given by
$$
\eqalign{
\Phi^{(1-i,i)}_\ve(u;u_0)
&=f(u-u_0)\bigoplus_{n,n'\atop n\in2\Z+m+i}
t_\ve(u-u_0)^{n'}_n \Phi(u)^{n'}_n,
\cr
\Phi^{(1-i,i)*}_\ve(u;u_0)
&=f^{-1}(u-u_0)\bigoplus_{n,n'\atop n\in2\Z+m+i}
t^*_\ve(u-u_0)^{n'}_n \Phi^*(u)^{n'}_n
=\Phi^{(1-i,i)}_{-\ve}(u-1)
}\eqlabel\EQvfvertex
$$
with the function
$$
\eqalign{
f(u)
&={1\over\sqrt{C}}x^{-u^2/2r+(r-1)u/2r+1/4} f_1(x^{2u}),
\cr
f_1(z)
&={1\over\sqrt{(x^{2r};x^{2r})_\infty}}
{(x^4z;x^4,x^{2r})_\infty(x^{2+2r}z^{-1};x^4,x^{2r})_\infty
\over(x^2z;x^4,x^{2r})_\infty(x^{2r}z^{-1};x^4,x^{2r})_\infty}.
}\eqlabel\EQffunction
$$
satisfying the equations
$$
C[u]\,f(u)f(u-1)=1,
\qquad {f(u-2)\over f(u)}={[u]\over[u-1]}.
\eqlabel\EQfequations
$$
These vertex operators satisfy Eqs.~(\EQevcommutation)
and (\EQevnormalization) and we want to identify them
with the operators $\Phi^{(1-i,i)}_\ve(u)$.
But they contain a new free parameter $u_0$, whereas
the final formulas for correlation functions like
(\EQevcorrfun) and (\EQevtrace) must be $u_0$ independent.
So let us look at the situation more carefully.
Any $u_0$-dependent operator
$O^{(i',i)}(u_0):\H_m^{(i)}\rightarrow\H_m^{(i')}$ with
$\H_m^{(i)}=\bigoplus_{n\in2\Z+m+i}\H_{m,n}$, which is defined in
the SOS model,
is related with the respective $u_0$-independent operator
$O^{(i',i)}:\H_i\rightarrow\H_{i'}$, which is defined in
the eight-vertex model, by the intertwining property
$$
O^{(i',i)}(u_0)U^{(i)}_m(u_0)=U^{(i')}_m(u_0)O^{(i',i)},
$$
where
$$
U^{(i)}_m(u_0)^{n_1n_2\ldots}_{\ve_1\ve_2\ldots}
=\bigoplus_n \hat t_{\ve_1}(u_0-u)^n_{n_1}
\hat t_{\ve_2}(u_0-u)^{n_1}_{n_2}\ldots
\quad:\quad
\H_m^{(i)}\rightarrow\H_i
$$
Here $u$ is the spectral parameter at the vertical arrows in Fig.~6c,
sequences $\ve_1,\ve_2,\ldots$ and $n,n_1,n_2,\ldots$ stabilize to
the respective ground state sequences, and each
intertwining vector is divided by its value on the ground
state configuration, which is denoted by hats.
So the traces in both cases must coincide
if the $u_0$-dependent operators are properly defined.

The `tail operator' $\Lambda(u)=\bigoplus_{n,n'}\Lambda(u)^{n'}_n$ is
$$
\Lambda(u){}^{n'\vphantom{n'_1}}_{n\vphantom{n_1}}
{}^{n'_1n'_2\ldots}_{n_1n_2\ldots}
=\L(n,n_1;n'_1,n'|u)\L(n_1,n_2;n'_2,n'_1|u)\ldots
\eqlabel\EQLambdadef
$$
with
$$
\L(n_1,n_2;n_3,n_4|u)
=\sum_\ve t^*_\ve(-u)^{n_2}_{n_1} t_\ve(-u)^{n_4}_{n_3}.
\eqlabel\EQLdef
$$
In fact, the r.~h.~s.\ of (\EQLambdadef) is a finite product because
all sequences $n_1,n_2,\ldots,n_k,\ldots$ stabilize to $(-)^{k+i}$
by definition.
Note also that we only need the elements with
$n'-n\in2\Z$ [or $n_1-n_4\in2\Z$
in Eq.~(\EQLdef)]. This is because only even number of vertex operators
enter into the correlation functions (see Fig.~6c).
So we shall imply it from here on.

The functions $L$ can be easily calculated explicitly:
$$
\eqalign{
\L(n,n\pm1;n'\pm1,n'|u)
&={[u\mp\half(n-n')][\half(n+n')]\over[u][n]},
\cr
\L(n,n\pm1;n'\mp1,n'|u)
&={[u\mp\half(n+n')][\half(n-n')]\over[u][n]}
}\qquad\hbox{for}\quad
n'-n\in2\Z.
\eqlabel\EQLexplicitform
$$
By definition
$$
\L(n,n';n'',n|u)=\delta_{n'n''},
\eqlabel\EQLdiag
$$
and hence
$$
\Lambda(u)^n_n=1.
\eqlabel\EQLambdadiag
$$

The counterpart of the operator $\rho^{(i)}$ is
$$
\rho^{(i)}(u_0)
=\const\bigoplus_{n',n\atop n\in2\Z+m+i}\Lambda(u_0)^{n'}_n[n]\,x^{4H_{m,n}}
\eqlabel\EQvfrho
$$

To prove the defining equations (\EQevkz)
we need only prove the properties (\EQevdiff-\EQevnormalization) of
the vertex operators. The commutation relation (\EQevcommutation) and the
normalization condition (\EQevnormalization) are proven
directly.\refs{\JMunpublished}%
\nfootnote{As well as we know, M.~Jimbo was the first
to make this observation.}
The most subtle point is the proof of Eq.~(\EQevdiff) for
$\Phi^{(1-i,i)}(u,u_0)$ and $\rho^{(i)}(u_0)$.
To do it we need the commutation relation between $\Phi_\ve(u;u_0)$ and
$\Lambda(u_0)$. As it shown in Fig.~8, we have the following relation
$$
\Lambda(u_0)^{n'}_{s}\Phi(u)^s_n
=\sum_{s'}\L(s,n;s',n'|u_0-u)
\Phi(u)^{n'}_{s'}\Lambda(u_0)^{s'}_n.
\eqlabel\EQLambdasosphicommutation
$$
Multiplying it by $f(u-u_0-2)t'_\ve(u-u_0)^s_n$,
taking sum in $s$, and using Eqs.~(\EQstarprimedef),
(\EQstarprimeexplicit) we have
$$
\Phi^{(1-i,i)}_\ve(u;u_0)\Lambda(u_0)
=\Lambda(u_0)f(u-u_0-2)\bigoplus_{n,n'}{[n']\over[n]}
t_\ve(u-u_0-2)^{n'}_n \Phi(u)^{n'}_n.
$$
We see that commutation with
$\Lambda$ shifts $u$ by $-2$ in each function except
$\Phi(u)^{n'}_n$ and adds the factor $[n']/[n]$. Now pulling
$\Phi(u)^{n'}_n$ through $[n]\,x^{4H_n}$ we obtain (\EQevdiff).

\topinsert
%
%
\line{\hss
\beginpicture
%
\setcoordinatesystem units <1cm,1cm> point at 0 0
\rl  1.0 3.0 , 4.2 3.0
\trl 1.5 2.0 , 4.2 2.0
\trl 1.0 1.0 , 4.2 1.0
\ln 0.507 2.0 , 1.007 1.0 \ln 0.493 2.0 , 0.993 1.0
\ln 1.007 3.0 , 1.507 2.0 \ln 0.993 3.0 , 1.493 2.0
\ln  2.0 3.0 , 2.5 2.0
\ln  3.0 3.0 , 3.5 2.0
\ln  4.0 3.0 , 4.2 2.6
\rl 0.7 1.6 , 0.8 1.6 \rl 0.8 1.6 , 0.8 1.4
\rl 1.2 2.6 , 1.2 2.4 \rl 1.2 2.4 , 1.3 2.4
\ln 0.8 1.6 , 1.2 2.4
\ln 1.4 1.0 , 1.55 1.1 \ln 1.55 1.1 , 1.6 1.0
\ln 1.9 2.0 , 1.95 1.9 \ln 1.95 1.9 , 2.1 2.0
\ln 1.55 1.1 , 1.95 1.9
\ln 2.4 1.0 , 2.55 1.1 \ln 2.55 1.1 , 2.6 1.0
\ln 2.9 2.0 , 2.95 1.9 \ln 2.95 1.9 , 3.1 2.0
\ln 2.55 1.1 , 2.95 1.9
\ln 3.4 1.0 , 3.55 1.1 \ln 3.55 1.1 , 3.6 1.0
\ln 3.9 2.0 , 3.95 1.9 \ln 3.95 1.9 , 4.1 2.0
\ln 3.55 1.1 , 3.95 1.9
\setdashes <2.5pt>
\rl 1.25 2.5 , 4.2 2.5 \ln 0.75 1.5 , 0.5 1.0
\rl 1.25 1.5 , 4.2 1.5 \ln 1.25 1.5 , 0.5 3.0
\rl 1.5 3.4 , 1.5 3.0 \ln 1.5 3.0 , 2.0 2.0 \rl 1.5 1.0 , 1.5 0.4
\setsolid
\ar 0.51 1.02 , 0.5 1.0
\ar 0.51 2.98 , 0.5 3.0
\ar 1.5 0.41 , 1.5 0.4
\put{$n$}  [t]  at 1.0 0.9
\put{$s$}  [r]  at 0.3 2.0
\put{$n'$} [B]  at 1.0 3.1
\put{$s'$} [t]  at 1.5 1.9
\put{$u$}   [t]  at 1.5 0.3
\put{$v$}   [rt] at 0.5 0.9
\put{$v_0$} [rb] at 0.5 3.1
\put{$=$} at 4.75 2.0
%
\setcoordinatesystem units <1cm,1cm> point at -5 0
\trl 1.0 3.0 , 4.2 3.0
\trl 1.0 1.0 , 4.2 1.0
\ln 0.507 2.0 , 1.007 1.0 \ln 0.493 2.0 , 0.993 1.0
\ln 1.4 3.0 , 1.5 2.9 \ln 1.5 2.9 , 1.6 3.0
\ln 1.4 1.0 , 1.5 1.1 \ln 1.5 1.1 , 1.6 1.0
\rl 1.5 2.9 , 1.5 1.1
\ln 2.4 3.0 , 2.5 2.9 \ln 2.5 2.9 , 2.6 3.0
\ln 2.4 1.0 , 2.5 1.1 \ln 2.5 1.1 , 2.6 1.0
\rl 2.5 2.9 , 2.5 1.1
\ln 3.4 3.0 , 3.5 2.9 \ln 3.5 2.9 , 3.6 3.0
\ln 3.4 1.0 , 3.5 1.1 \ln 3.5 1.1 , 3.6 1.0
\rl 3.5 2.9 , 3.5 1.1
\rl 0.7 1.6 , 0.8 1.6 \rl 0.8 1.6 , 0.8 1.4
\ln 0.8 1.6 , 1.125 2.25 \rl 1.125 2.25 , 4.2 2.25
\setdashes <2.5pt>
\ln 0.75 1.5 , 0.5 1.0
\ln 1.125 1.75 , 0.5 3.0 \rl 1.125 1.75 , 4.2 1.75
\rl 1.5 3.4 , 1.5 3.0 \rl 1.5 1.0 , 1.5 0.4
\setsolid
\ar 0.51 1.02 , 0.5 1.0
\ar 0.51 2.98 , 0.5 3.0
\ar 1.5 0.41 , 1.5 0.4
\put{$n$}  [t]  at 1.0 0.9
\put{$s$}  [r]  at 0.3 2.0
\put{$n'$} [B]  at 1.0 3.1
\put{$u$}   [t]  at 1.5 0.3
\put{$v$}   [rt] at 0.5 0.9
\put{$v_0$} [rb] at 0.5 3.1
\put{$=$} at 4.75 2.0
%
\setcoordinatesystem units <1cm,1cm> point at -10 0
\trl 1.0 3.0 , 4.2 3.0
\trl 0.5 2.0 , 4.2 2.0
\rl  1.0 1.0 , 4.2 1.0
\ln 1.4 3.0 , 1.45 2.9 \ln 1.45 2.9 , 1.6 3.0
\ln 0.9 2.0 , 1.05 2.1 \ln 1.05 2.1 , 1.1 2.0
\ln 1.45 2.9 , 1.05 2.1
\ln 2.4 3.0 , 2.45 2.9 \ln 2.45 2.9 , 2.6 3.0
\ln 1.9 2.0 , 2.05 2.1 \ln 2.05 2.1 , 2.1 2.0
\ln 2.45 2.9 , 2.05 2.1
\ln 3.4 3.0 , 3.45 2.9 \ln 3.45 2.9 , 3.6 3.0
\ln 2.9 2.0 , 3.05 2.1 \ln 3.05 2.1 , 3.1 2.0
\ln 3.45 2.9 , 3.05 2.1
\ln 3.9 2.0 , 4.05 2.1 \ln 4.05 2.1 , 4.1 2.0
\ln 4.05 2.1 , 4.2 2.4
\ln 0.5 2.0 , 1.0 1.0
\ln 1.5 2.0 , 2.0 1.0
\ln 2.5 2.0 , 3.0 1.0
\ln 3.5 2.0 , 4.0 1.0
\setdashes <2.5pt>
\rl 0.3 2.5 , 4.2 2.5
\rl 0.3 1.5 , 4.2 1.5
\rl 1.5 3.4 , 1.5 3.0 \ln 1.0 2.0 , 1.5 1.0 \rl 1.5 1.0 , 1.5 0.4
\setsolid
\ar 0.31 2.5 , 0.3 2.5
\ar 0.31 1.5 , 0.3 1.5
\ar 1.5 0.41 , 1.5 0.4
\put{$n$}  [t]  at 1.0 0.9
\put{$s$}  [r]  at 0.3 2.0
\put{$n'$} [B]  at 1.0 3.1
\put{$u$} [t]  at 1.5 0.3
\put{$v$} [r]  at 0.2 1.5
\put{$v\smash{{}_0}$} [r] at 0.2 2.5
\endpicture
\hss}
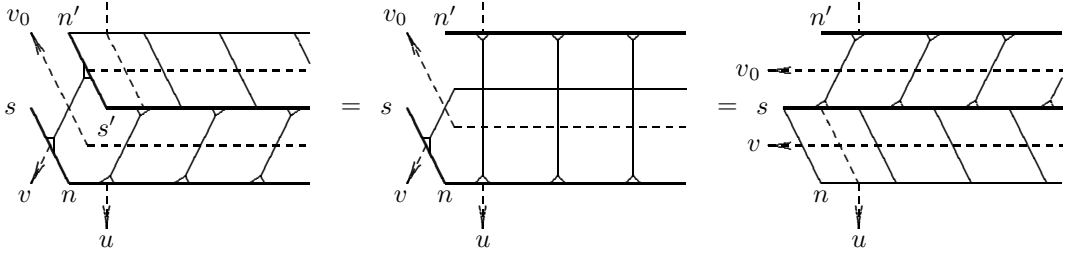
\figcap{8} Commutation of $\Phi(u-v)^{n'}_n$ with $\Lambda(u_0-u)$. Summation
in $s'$ is implied.\par
\endinsert

It is necessary also to ensure that the spectrum of $\rho^{(i)}(u_0)$
from (\EQvfrho) coincides with the spectrum of $\rho^{(i)}$ from
(\EQevrho). To do it let us calculate the quantity
$$
\chi^{\prime(i)}_m(x^4)
=\sum_{n\in2\Z+m+i}\Tr_{\H_{m,n}}(\Lambda(u_0)^n_n[n]\,x^{4H_{m,n}})
=\sum_{n\in2\Z+m+i}[n]\,\chi_{m,n}(x^4).
\eqlabel\EQprimecharacter
$$
In the Appendix B it is shown that
$$
\chi^{\prime(i)}_m(x^4)=[m]'\chi^{(i)}(x^4)
\eqlabel\EQchicheck
$$
with
$$
[u]'=x^{{u^2\over r-1}-u}\Theta_{x^{2(r-1)}}(x^{2u})
=\sqrt{\pi\over\epsilon(r-1)}\,\e^{\quarter\epsilon(r-1)}\,
\theta_1\!\left({u\over r-1};{\i\pi\over\epsilon(r-1)}\right).
$$
We expect that traces of operators (\EQvfvertex) and (\EQvfrho)
are proportional to $[m]'$ so that this
factor cancels in the probabilities. This expectation is realized
at least in the case of two vertex operators in the trace (see Appendix D).
In other words, we expect that
$$
\Lambda(u_0)\rho_m=\const[m]'x^{4H^{(i)}(u_0)}
$$
with $H^{(i)}(u_0)$ being an operator with the same spectrum
as the eight-vertex corner Hamiltonian $H^{(i)}$ and
$\rho_m=\bigoplus_n\rho_{m,n}$.

\nsec Free Field Representation for the Eight-Vertex Model\negbigskip

\ssubsec{1} Free Field Representation for the SOS model\refs{\LP}

The most efficient way to calculate correlation functions
in the SOS model is based on the free field representation.
Consider a Heisenberg algebra of operators $a_n$ with
nonzero integer $n$ and a pair of `zero mode' operators $\P$ and $\Q$
with the commutation relations
$$
[\P,\Q]=-\i,\qquad
[a_k,a_l]=k{[k]_x[(r-1)k]_x\over[2k]_x[rk]_x}\delta_{k+l,0}
\quad\hbox{with}\quad
[u]_x={x^u-x^{-u}\over x-x^{-1}}.
\eqlabel\EQheisenberg
$$
The normal ordering operation $\lcolon\ldots\rcolon$
places $\P$ to the right of $\Q$ and $a_k$ with positive $k$
to the right of $a_{-k}$. Now introduce the field
$$
\varphi(z)=-\sqrt{r-1\over2r}(\Q-\i\P\log z)
-\sum_{k\neq0}{a_k\over\i k}z^{-k}.
\eqlabel\EQvarphidef
$$
This field enters into the exponential operators
$$
V(u)=z^{(r-1)/4r}\lcolon\e^{\i\varphi(z)}\rcolon,
\qquad
\bar V(u)=z^{(r-1)/r}\lcolon\e^{-\i\varphi(x^{-1}z)-\i\varphi(xz)}\rcolon,
\eqlabel\EQVbarVdef
$$
and Lukyanov's screening operator
$$
\eqalign{
X(u)
&=\eta^{-1}\epsilon\int_{C_u}{dv\over\i\pi}\,
\bar V(v){[v-u+\half-\sqrt{2r(r-1)}\,\P]
\over[v-u-\half]},
\cr
\eta^{-1}
&=\i[1]\,x^{r-1\over2r}
{(x^2;x^{2r})_\infty\over(x^{2r-2};x^{2r})_\infty}
{(x^6;x^4,x^{2r})_\infty(x^{2r+2};x^4,x^{2r})_\infty
\over
(x^4;x^4,x^{2r})_\infty(x^{2r+4};x^4,x^{2r})_\infty}.
}\eqlabel\EQscreening
$$
The contour $C_u$ goes from $u$ to $u+{\i\pi\over\epsilon}$.

Now the vertex operators are represented as follows:
$$
\eqalign{
\Phi(u)^{n+1}_n
&={\i^{m-n}\over[n]}V(u),
\cr
\Phi(u)^{n-1}_n
&=(-)^{m-n+1}{\i^{m-n}\over[n]}V(u)X(u).
}\eqlabel\EQbosvertex
$$
The corner Hamiltonian is given by
$$
H={\P^2\over2}+\sum_{k=1}^\infty
{[k]_x[rk]_x\over[2k]_x[(r-1)k]_x}a_{-k}a_k.
\eqlabel\EQbosH
$$
The operators (\EQbosvertex) and (\EQbosH) act in the direct
sum of the Fock spaces $\F_{m,n}$
generated by the operators $a_{-k}$ with $k>0$ from the highest
weight vectors $|P_{m,n}\rangle$ such that
$$
a_k|P_{m,n}\rangle=0\quad (k>0),
\qquad
\P|P_{m,n}\rangle=P_{m,n}|P_{m,n}\rangle,
\qquad
P_{m,n}=m\,\sqrt{r\over2(r-1)}-n\,\sqrt{r-1\over2r}.
\eqlabel\EQPvacdef
$$
It is not clear if the spaces $\bigoplus_n\H_{m,n}$ and
$\bigoplus_n\F_{m,n}$ coincide as representations of the
vertex operator algebra.
At least the numbers of states at each level (according to the
$H$ grading) of both spaces $\H_{m,n}$ and $\F_{m,n}$ coincide.
We only assume that traces over both spaces coincide.

Later we shall need the following fact on the SOS model.
The weights (\EQWmatrix) are invariant under the substitution
$n_k\to -n_k$. So the model is invariant with respect to the change
$$
\eqalign{
n_k&\to-n_k
\cr
m&\to-m.
}\eqlabel\EQchangesigns
$$
The free field representation (\EQbosvertex) is not invariant under this
substitution. So there is another free field representation
$$
\eqalign{
\Phi(u)^{n+1}_n
&={\i^{m-n}\over[n]}V(u)X(u),
\cr
\Phi(u)^{n-1}_n
&=(-)^{m-n+1}{\i^{m-n}\over[n]}V(u).
}\eqlabel\EQbosvertexsecond
$$
In this representation the space $\H_{m,n}$ is identified with
$\F_{-m,-n}$. Of course, this representation gives the same
local heights probabilities as the first representation.

\ssubsec{2} Free Field Representation for the Operator $\Lambda(u)$

Now let us turn to the eight-vertex model. The representation
of $\rho^{\prime(i)}$ and $\Phi^{(1-i,i)}_\ve(u;u_0)$ completely reduces
to the free field representation of the SOS model by use of
Eqs.~(\EQvfcorner) and (\EQvfvertex). The only unknown
object is $\Lambda(u_0)$. To find its representation let us consider
the limit $u_0\to u$ in the commutation relation (\EQLambdasosphicommutation).
In this limit $\L(s,n;s',n'|u)\to\infty$ and
$$
\L(n\pm1,n;n'+1,n'|u)\bigg/\L(n\pm1,n;n'-1,n'|u)\to1,
\qquad u\to0.
$$
We have
$$
\Phi(u)^{n'}_{n'-1}\Lambda(u)^{n'-1}_n
=-\Phi(u)^{n'}_{n'+1}\Lambda(u)^{n'+1}_n.
\eqlabel\EQLambdasosphiequalu
$$
Substitution of (\EQbosvertex) into (\EQLambdasosphiequalu) gives
$$
V(u)\Lambda(u)^{n'-1}_n
=(-)^{m-n'}{[n'-1]\over[n'+1]}V(u)X(u)\Lambda(u)^{n'+1}_n.
$$
For $n'<n$ and odd $n'-n$ we have a solution with the initial
condition (\EQLambdadiag):
$$
\Lambda(u)^{n-2k}_n
=(-)^{(m-n+1)k}{[n-2k]\over[n]}X^k(u)\quad\hbox{for}\quad k\geq0.
\eqlabel\EQbosLambda
$$
In the Appendix C we prove that this solution satisfies
the general commutation relation (\EQLambdasosphicommutation).
So we may consider Eq.~(\EQbosLambda) as the free field representation
of $\Lambda$. Now we have to construct representatives for
$\Lambda(u)^{n'}_n$ with $n'>n$. However, we know no such
representatives in the representation (\EQbosvertex).
To solve this problem let us use the second
free field representation of the SOS model (\EQbosvertexsecond),
described at the end of Sec.~3. In this representation there is
a natural representative
$$
\Lambda(u)^{n+2k}_n=(-)^{(m-n+1)k}{[n+2k]\over[n]}X^k(u),
\quad\hbox{for}\quad k\geq0,
\eqlabel\EQbosLambdasecond
$$
acting from $\F_{-m,-n}$ to $\F_{-m,-n-2k}$. So we must
use the representation (\EQbosvertex), (\EQbosLambda) for
the contributions with $n'\leq n$, and the representation
(\EQbosvertexsecond), (\EQbosLambdasecond) for those with $n'\geq n$.

\ssubsec{3} Staggered Spontaneous Polarization

Let us check our construction in the calculation of the staggered
polarization in the antiferroelectric regime of the eight-vertex
model. The staggered polarization is the average `spin'
$\langle\ve\rangle^{(i)}$ in the state with the condition at
the infinity $i$. According to (\EQevcorrfun) it is given by
$$
\langle\ve\rangle^{(i)}
=\sum_\ve \ve P_\ve^{(i)}
={1\over\Tr\rho^{(i)}(u_0)}
\sum_\ve\ve\Tr\left(
\Phi^{(i,1-i)}_{-\ve}(u-1,u_0)\,\Phi^{(1-i,i)}_\ve(u,u_0)
\,\rho^{(i)}(u_0)\right).
\eqlabel\EQstaggereddef
$$
Substituting (\EQvfvertex) we obtain
$$
\eqalignno{
\langle\ve\rangle^{(i)}
&={(-1)^i\over[m]'}(x^2;x^4)_\infty
\sum_{n_2\in{\bf2Z}+m+i}\sum_{n_1=n_2\pm1\atop n_0=n_1\pm1}
[n_2]\,\K(n_1,n_2;n_0,n_1|u-u_0)
\cr
&\quad\times
\Tr_{\F_{m,n_2}}\left(
\Phi(u-1)^{n_2}_{n_1}\Phi(u)^{n_1}_{n_0}
\Lambda(u_0)^{n_0}_{n_2}x^{4H_n}\right),
\lnlabel\EQstaggeredrepr}
$$
where
$$
(-)^{n_3-m}\K(n_1,n_2;n_3,n_4|u)
=f(u)f(u-1)\sum_\ve \ve\,t_{-\ve}(u-1)^{n_2}_{n_1} t_\ve(u)^{n_4}_{n_3}.
\eqlabel\EQKdef
$$
Applying the trick with the sign change of $m$ and $n$s we obtain
$$
\eqalignno{
\langle\ve\rangle^{(i)}
&={(-1)^i\over[m]'}(x^2;x^4)_\infty
\sum_{n\in2\Z+m+i}[n]
\left(K^{+-}_n(T^{+-}_{m,n}-T^{+-}_{-m,n})
+K^{++}_n(T^{++}_{m,n}-T^{++}_{-m,n})\right)
\lnlabel\EQstaggeredtrick
\cr
\span
K^{+-}_n=\K(n-1,n;n,n-1|u-u_0),
\qquad K^{++}_n=\K(n-1,n;n-2,n-1|u-u_0),
\cr
\span
\eqalign{
T^{+-}_{m,n}
&=\Tr_{\F_{m,n}}(\Phi(u-1)^n_{n-1}\Phi(u)^{n-1}_nx^{4H_n}),
\cr
T^{++}_{m,n}
&=\Tr_{\F_{m,n}}(\Phi(u-1)^n_{n-1}\Phi(u)^{n-1}_{n-2}
\Lambda(u_0)^{n-2}_nx^{4H_n}).
}
}
$$
In Appendix D we perform the direct calculation using the free field
representation. The answer is $m$ and $u_0$ independent and is
given by the well-known Baxter--Kelland formula\refs{\BK}
$$
\langle\ve\rangle^{(i)}
=-(-1)^i{(x^2;x^2)_\infty^2\over(-x^2;x^2)_\infty^2}
\int_{C_0}{\epsilon dv\over\i\pi}\,
{\theta_4\!\left({v-1/2\over r};{\i\pi\over\epsilon r}\right)
\over\theta_1\!\left({v-1/2\over r};{\i\pi\over\epsilon r}\right)}
=(-)^i{(x^2;x^2)_\infty^2(-x^{2r};x^{2r})_\infty^2
\over(-x^2;x^2)_\infty^2(x^{2r};x^{2r})_\infty^2}.
\eqlabel\EQBK
$$
This gives an additional `finite temperature' check of the correspondence
of ground states of the eight-vertex and SOS models.

\nsec Discussion

We proposed a free field construction for correlation functions
of the eight-vertex model. The main scheme is the following.
We have the bosonic representation for the SOS model, and we know
the vertex-face correspondence. Using it we can express
objects of the eight-vertex models in terms of the objects
in the SOS model. The only nontrivial point is the representation
of the operator $\Lambda(u)$ --- the unremovable `tail' of the intertwining
vectors. The matrix elements $\Lambda(u)^{n-2k}_n$ with $k\geq0$
are represented up to a constant factor as powers of
Lukyanov's screening operator $X^k(u)$, and $\Lambda(u)^{n+2k}_n$ remain
undefined. Nevertheless, it is possible to calculate necessary
correlation functions using the fact that the SOS model
admits of two free field representations.

\goodbreak
Our construction is checked in several ways:

\nobreak
\ $\bullet$~it supplies solutions to the defining relations for correlation
functions;

\ $\bullet$~it gives the correct spectrum of the corner Hamiltonian;

\ $\bullet$~it gives the correct average staggered polarization.

Nevertheless there still remain some open problems. The most fundamental
problem is why we may substitute traces over the configuration space
$\H_{m,n}$ by traces over the Fock $\F_{m,n}$? In the case of the
six-vertex model the identification of the configuration and Fock
spaces is based on the quantum affine symmetry and low temperature
expansions\refs{\DFJMN,\KashMiwa}.
In the case of SOS model there is no such symmetry algebra, which could
be defined directly on the lattice. We believe that the solution of
this problem could also clarify inconsistency of the involutions in the
algebra of vertex operators and in the Heisenberg algebra.
This inconsistency forces us to use two free field representations
for vertex operators to find correlation functions.

An important problem is
to prove $m$ and $u_0$ independence of all correlation functions in
the framework of the free field representation. This would be a strong
argument for the consistency of our approach.

Another group of questions is related with the problem of calculation
of form factors. For the six-vertex, SOS, and RSOS models we know that
there exist type II vertex operators, that describe asymptotic states
in the models, and allow one to obtain general
form factors of local operators. The existence of such vertex
operators is substantiated by the construction of the elliptic
algebra\refs{\JMellipticalg}. To find the representation for
the form factors it seems to be necessary to find a free field
representation of generic matrix elements $\Lambda(u_0)^{m'}_m{}^{n'}_n$.

\sec* Acknowledgments

We are grateful to A.~Antonov, A.~Belavin, Vl.~Dotsenko,
M.~Jimbo, S.~Lukyanov, and T.~Miwa for discussions.
We are especially indebted to M.~Jimbo for sending us his unpublished
notes on the problem. The work was supported in part by the CRDF under
the grant RP1--277, and by INTAS and RFBR under the grant
INTAS--RFBR--95--690.

\secno=-1
\sec* Appendix A

Let us consider a trace like (\EQstaggereddef). The formal series that define
this trace is convergent in the region
$$
|x^{-4}z_0|>|x^{-2}z|>|z|>|z_0|
\quad\hbox{or}\quad
\Re u_0-2<\Re u-1<\Re u<\Re u_0.
$$
So we must consider the trace for arbitrary values of $u$ and $u_0$
as an analytic continuation from this region.
Besides we know that in the physical region (the regime $III$)
$$
0<u<1.
$$

As a consequence of these conditions we have
$$
-1<\Re(u-u_0)<0,\qquad -2<\Re(-u_0)<0.
$$
On the other hand the intertwining vectors are contained in three
objects: operator $\Phi_\ve(u,u_0)$ contains $t_\ve(u-u_0)$;
operator $\Phi_\ve(u-1,u_0)$ contains $t^*_\ve(u-u_0)$;
operator $\rho(u_0)$ contains $t_\ve(-u_0)$ and $t^*_\ve(-u_0)$.
So all the intertwining
vectors $t_\ve(v)^{n'}_n$, $t^*_\ve(v)^{n'}_n$ entering into
the correlation function representation
are of the argument in the region
$$
-2<\Re v<0.
$$

\secadvance
\sec* Appendix B

Here we prove the identity
$$
\sum_{n\in2\Z+m+i}[n]\,\chi_{m,n}(x^4)=[m]'\chi^{(i)}(x^4).
\eqlabel\EQcharidentity
$$
We have
$$
\eqalignno{
\sigma^{(i)}
&=\sum_{n\in2\Z+m+i}[n]\,\chi_{m,n}(x^4)
\cr
&=\sum_{n\in2\Z+m+i}[n]\,{x^{2P_{m,n}^2}\over(x^4;x^4)_\infty}
=\sum_{n\in2\Z+m+i}{x^{4h'_{m,n}}\over(x^4;x^4)_\infty}
\Theta_{x^{2r}}(x^{2n})
}
$$
with
$$
h'_{m,n}={P^2_{m,n}\over2}+\quarter\left({n^2\over r}-n\right)
={n^2\over4}-{(2m+1)n\over4}+\quarter{r\over r-1}m^2.
$$
We see that the fractional coefficient at $n^2$ in $2P^2_{m,n}$
is compensated in $4h'_{m,n}$ by the fractional coefficient
that takes its origin in $[n]$.

Now let us apply the identity
$$
\Theta_p(z)=\sum_{k\in\Z}(-1)^k p^{k(k+1)/2} z^{-k}.
$$
We obtain
$$
\eqalignno{
\sigma^{(i)}
&={1\over(x^4;x^4)_\infty}
\sum_{k\in\Z}(-1)^k x^{rk(k+1)+{r\over r-1}m^2}a^{(i)}_k,
\cr
a^{(i)}_k
&=\sum_{n\in2\Z+m+i}x^{n^2-(2m+2k+1)n}
\cr
&=x^{-(k+m)(k+m+1)}\sum_{l\in\Z+\half(k+i)}x^{4l^2-2l}
=x^{-(k+m)(k+m+1)}\sum_{l\in\Z}x^{4l^2\mp2l}
\cr
&=x^{-(k+m)(k+m+1)}\Theta_{x^8}(-x^2)
=x^{-(k+m)(k+m+1)}{(x^4;x^4)_\infty\over(x^2;x^4)_\infty}.
}
$$
Here we used the substitution $n\to 2l+k+m$ and then $l\to l+\half$ for
odd $k$. The equality in the last line is easily obtained by use of
the evident identities
$$
(z;p)_\infty=(z;p^2)_\infty(zp;p^2)_\infty,
\qquad
(z^2;p^2)_\infty=(z;p)_\infty(-z;p)_\infty.
$$
So we have
$$
\sigma^{(i)}
={1\over(x^2;x^4)_\infty}
\sum_{k\in\Z}(-1)^k x^{(r-1)k(k+1)-2mk+{m^2\over r-1}-m}
={[m]'\over(x^2;x^4)_\infty},
$$
q.~e.~d.

Note that this proof, as well as the consideration in Appendix D,
is valid for $m,n\in\Z+\delta$ with any $\delta$.

\secadvance
\sec* Appendix C

In this Appendix we prove that the operators (\EQbosLambda) satisfy
the commutation relation (\EQLambdasosphicommutation). Namely, consider the
quantity
$$
\eqalignno{
d(u,u_0)
&=\Lambda(u_0)^{n+1-2k}_{n+1}\Phi(u)^{n+1}_n
-\L(n+1,n;n+2-2k,n+1-2k|u_0-u)
\Phi(u)^{n+1-2k}_{n+2-2k}\Lambda(u_0)^{n+2-2k}_n
\cr
&\quad
-\L(n+1,n;n-2k,n+1-2k|u_0-u)\Phi(u)^{n+1-2k}_{n-2k}\Lambda(u_0)^{n-2k}_n
\lnlabel\EQCddef}
$$
We have to prove that $d(u,u_0)=0$ from the free field representation
inputs. Substituting Eq.~(\EQbosvertex) we obtain
$$
\eqalign{
d(u,u_0)
&\sim[n+1-2k]X^k(u_0)V(u)
+(-1)^k{[u_0-u+n+1-k][k]\over[u_0-u]}V(u)X(u)X^{k-1}(u_0)
\cr
&\quad
-(-1)^k{[u_0-u+k][n+1-k]\over[u_0-u]}V(u)X^k(u_0).
}
$$
Here the sign $\sim$ only means that two functions differ
by a factor which is allowed to be any nonzero function of all variables.
Substituting Eq.~(\EQscreening) and taking into account that
$$
V(u_1)\bar V(u_2)=-{[u_1-u_2+\half]\over[u_1-u_2-\half]}\bar V(u_2)V(u_1)
\eqlabel\EQCVXcommutation
$$
we obtain
$$
d(u,u_0)
\sim\int{d^k v\over(2\pi\i)^k}
V(u)\bar V(v_k)\ldots\bar V(v_1)f(u,u_0;v_1,\ldots,v_k)
$$
with
$$
\eqalignno{
f(u,u_0;v_1,\ldots,v_k)
&=\Sym\biggl\{
[n+1-2k]
\prod_{i=1}^k{[v_i-u_0+\half-n-1+2(i-1)]\over[v_i-u_0-\half]}
{[v_i-u+\half]\over[v_i-u-\half]}
\cr
&\quad
+{[u_0-u+n+1-k][k]\over[u_0-u]}
{[v_k-u-n+2k-{3\over2}]\over[v_k-u-\half]}
\prod_{i=1}^{k-1}{[v_i-u_0+\half-n+2(i-1)]\over[v_i-u_0-\half]}
\cr
&\quad
-{[u_0-u+k][n+1-k]\over[u_0-u]}
\prod_{i=1}^k{[v_i-u_0+\half-n+2(i-1)]\over[v_i-u_0-\half]}
\biggr\}.
\lnlabel\EQCfdef}
$$
Here the operation $\Sym$ is defined as
$$
\Sym F(v_1,\ldots,v_k)={1\over k!}
\sum_{\sigma\in S_k} F(v_{\sigma(1)},\ldots,v_{\sigma(k)})
\prod_{i<j\atop\sigma(i)>\sigma(j)}
h(v_{\sigma(i)}-v_{\sigma(j)})
$$
with $h(v)$ being defined as
$$
\bar V(v_1)\bar V(v_2)=h(v_1-v_2)\bar V(v_2)\bar V(v_1),
\qquad h(v)={[v-1]\over[v+1]}.
$$
This `symmetrization' of the function $f$
corresponds to symmetrization of the integrand.

So we have to prove that $f(u,u_0;v_1,\ldots,v_k)=0$.

To calculate $f$ let us use the identity\refs{\JLMP}
$$
\Sym\prod_{i=1}^k [v_i-2i+2]
={[k]!\over k![1]^k}
\prod_{i<j}{[v_i-v_j]\over[v_i-v_j-1]}\prod_{i=1}^k[v_i-k+1]
$$
with $[k]!=\prod_{i=1}^k[i]$. We apply this function directly
to the first and last terms, but in the second term we only apply
it to $k-1$ variables, and symmetrize in the remaining variable
by hand. We have
$$
f(u,u_0;v_1,\ldots,v_N)\sim g(u)
$$
with $g(u)$ being a doubly periodic function of $u$ (and all
other variables):
$$
\eqalignno{
g(u)
&=1-{[n+1-2k]\over[n-k+1]}{[u_0-u]\over[u_0-u+k]}
\prod_{i=1}^k
{[v_i-u_0-n+k-{3\over2}][v_i-u+\half]
\over[v_i-u_0-n+k-\half][v_i-u-\half]}
\cr
&\quad
-{[u_0-u+n+1-k][1]\over[u_0-u+k][n-k+1]}
\prod_{i=1}^k{[v_i-u_0-n+k-{3\over2}]\over[v_i-u_0-n+k-\half]}
\cr
&\quad\times
\sum_{j=1}^k\left({[v_j-u-n+2k-{3\over2}][v_j-u_0-\half]
\over[v_j-u-\half][v_j-u_0-n+k-{3\over2}]}
\prod_{i\neq j}{[v_i-v_j+1]\over[v_i-v_j]}\right).
}
$$
It is easy to check that
$$
g(u_0+n+1-k)=0.
$$
So it is enough to prove that the function $g(u_0)$ is constant.
It has poles at the points (up to periods)
$$
\eqalign{
u&=u_0+k,
\cr
u&=v_l-1/2.
}
$$
It is straightforward to check that the residue at $u=v_l-\half$
vanishes. For the residue at $u=u_0+k$ we have
$$
\Res_{u=u_0+k} g(u)
\sim g_0(u_0)=1-{1\over[k]}
\prod_{i=1}^k{[v_i-u_0-k-\half]\over[v_i-u_0-k+\half]}
\sum_{j=1}^k\left({[v_j-u_0-\half]\over[v_j-u_0-k-\half]}
\prod_{i\neq j}{[v_i-v_j+1]\over[v_i-v_j]}\right).
$$
It is easy to check that $g_0(v_l-k-\half)=0$. For the residues
we have
$$
\Res_{u_0=v_{l(1)}-k+\half} g_0(u_0)\sim g_1(v_{l(1)})
=1-{1\over[k-1]}\sum_{j\neq l(1)}
\left({[v_j-v_{l(1)}+k-1]\over[v_j-v_{l(1)}+1]}
\prod_{i\neq j,l(1)}
{[v_j-v_i-1][v_{l(1)}-v_i]\over[v_j-v_i][v_{l(1)}-v_i-1]}\right).
$$
To proceed by induction introduce the functions
$$
g_s(v_{l(s)})=1-{1\over[k-s]}
\sum_{j\neq l(1),\ldots,l(s)}
\left(
\prod_{p=1}^s{[v_j-v_{l(p)}+k-1]\over[v_j-v_{l(p)}+1]}
\prod_{i\neq j,l(1),\ldots,l(s)}
\prod_{p=1}^s{[v_j-v_i-1][v_{l(p)}-v_i]\over[v_j-v_i][v_{l(p)}-v_i-1]}
\right).
$$
We suppose that $l(1),\ldots,l(k)$ is an arbitrary permutation
of the numbers $1,\ldots,k$ (i.~e.\ $l\in S_k$). We have
$$
g_s(v_{l(s+1)})=0,
\qquad
\Res_{v_{l(s)}=v_{l(s+1)}+1}g_s(v_{l(s)})\sim g_{s+1}(v_{l(s+1)}),
\qquad g_k(v_{l(k)})=0.
$$
Hence it follows that all $g_s$ and $f$ vanish.
This completes the proof in the case $s=n+1$. The proof in the case $s=n-1$
is similar though more cumbersome.

\secadvance
\sec* Appendix D

Here we show that the free field representation for the eight-vertex
model results in the Baxter--Kelland formula for
the staggered polarization.

As the initial point of the calculation we take Eq.~(\EQstaggeredtrick).
For the coefficients $K$ we have
$$
\eqalign{
\K(n,n\pm1;n'\pm1,n'|u)
&={h_4(u\pm{n-n'\over2})\,h_4({n+n'\over2})\over h_1(u)},
\cr
\K(n,n\pm1;n'\mp1,n'|u)
&={h_4(u\pm{n+n'\over2})\,h_4({n-n'\over2})\over h_1(u)}
}\eqlabel\EQDKcoeffs
$$
for $n'-n\in2\Z$, with
$$
h_i(u)=\sqrt{\pi\over\epsilon r}\,\e^{\quarter\epsilon r}
\theta_i\!\left({u\over r};{\i\pi\over\epsilon r}\right),
\qquad h_1(u)=[u],
\qquad
h_4(u)=\i\e^{-{\pi^2\over4r\epsilon}-\i\pi{u\over r}}[u-\i\pi/2\epsilon\,].
$$
Hence
$$
\eqalign{
K^{+-}_n
&={h_4(n-1)h_4(u)\over h_1(u)},
\cr
K^{++}_n
&={h_4(n-1+u)h_4(0)\over h_1(u)}.
}\eqlabel\EQDKsmallcoeffs
$$
The traces $T^{+-}_{m,n}$ and $T^{++}_{m,n}$ can be expressed
as follows
$$
T^{+-}_{m,n}=-{\i\over[n-1][n]}T_{m,n}(u,u),
\qquad
T^{++}_{m,n}={\i\over[n-1][n]}T_{m,n}(u,u_0)
$$
with
$$
\eqalignno{
T_{m,n}(u,u_0)
&=\Tr_{\F_{m,n}}(V(u-1)V(u)X(u_0)x^{4H})
\cr
&=\epsilon\int{dv\over\i\pi}H_{m,n}(u,u_0;v)G(v-u).
\lnlabel\EQDHGdef}
$$
Here $G(v)$ is the contribution into the trace in the second line
from the oscillators $a_n$:
$$
\eqadvance\EQDGdef
\eqalignno{
G(v)
&={\eta^{-1}\N\over(x^4;x^4)_\infty}
{(x^2;x^2)_\infty\over(x^{2r};x^{2r})_\infty}
{\Theta_{x^{2r}}(x^{-1}\zeta)\over\Theta_{x^2}(x^{-1}\zeta)}
\cr
&=\i{(x^2;x^2)_\infty^2\over(-x^2;x^2)_\infty}
\e^{\epsilon{r-1\over2r}-\epsilon{r-1\over r}(v-\half)^2}
{h_1(v-\half)
\over\vartheta_1(v-\half)},
\qquad\zeta=x^{2v},
\lnlabelno(\EQDGdef a)
\cr
\eta^{-1}\N
&=\i x^{-{r-1\over2r}}(x^2;x^2)_\infty^2(x^{2r};x^{2r})_\infty,
\qquad
\vartheta_i(u)=\sqrt{\pi\over\epsilon}\,\e^{\quarter\epsilon}
\theta_i(u;\i\pi/\epsilon).
\lnlabelno(\EQDGdef b)}
$$
The factor $\N$ originates from the traces of individual exponential
operators, whereas the rest originates from traces of pair products
in the generalized Wick theorem.

The factor $H_{m,n}(u,u_0;v)$ denotes the contribution of the `zero mode':
$$
\eqalignno{
H_{m,n}(u,u_0;v)
&=\Bigl\langle P_{m,n}\,\Bigl|\,
x^{2\P^2}
\e^{-\i\sqrt{r-1\over2r}\Q}\,(x^{-2}z)^{-\sqrt{r-1\over2r}\P+{r-1\over4r}}\,
\e^{-\i\sqrt{r-1\over2r}\Q}\,z^{-\sqrt{r-1\over2r}\P+{r-1\over4r}}
\cr
&\quad\times
\e^{\i\sqrt{2{r-1\over r}}\Q}\,\zeta^{\sqrt{2{r-1\over r}}\P+{r-1\over r}}
\,{[v-u_0+1/2-\sqrt{2r(r-1)}\P]\over[v-u_0-1/2]}
\,\Bigr|\,P_{m,n}\Bigr\rangle
\cr
&=(-)^{n-m}\e^{-2\epsilon{r-1\over r}(v-u+{1\over4})
-2\epsilon(v-u+{1\over2})m-\epsilon{r\over r-1}m^2
-\epsilon{r-1\over r}n^2+2\epsilon mn+2\epsilon(v-u+{1\over2})n}
\cr
&\quad\times
{h_1(v-u_0+\half-n)\over h_1(v-u_0-\half)}.
\lnlabel\EQDHdef}
$$
It is convenient to calculate the quantity
$$
\eqalignno{
H^{(i)}_m(u,u_0;v)
\hskip-2em
&\hskip2em
=\sum_{n\in2\Z+m+i}{1\over[n-1]}
(K_n^{++}H_{m,n}(u,u_0;v)-K_n^{+-}H_{m,n}(u,u;v))
\cr
&=(-)^i\sum_n{\e^{-2\epsilon{r-1\over r}(v-u+{1\over4})
-2\epsilon(v-u+{1\over2})m-\epsilon{r\over r-1}m^2
-\epsilon{r-1\over r}n^2-2\epsilon mn
+2\epsilon{r-1\over r}(v-u+{1\over2})n}
\over
h_1(u-u_0)\,h_1(n-1)\,h_1(v-u-\half)\,h_1(v-u_0-\half)}
\cr
&\quad\times
\textstyle
\bigl\{
h_1(v-u-\half)\,h_1(v-u_0+\half-n)\,h_4(n-1+u-u_0)\,h_4(0)
\cr
&\quad
\textstyle
-h_1(v-u_0-\half)\,h_1(v-u+\half-n)\,h_4(n-1)\,h_4(u-u_0)
\bigr\}.
\lnlabel\EQDHcomb}
$$
Substituting the identity
$$
\eqalignno{
&\theta_4(2x)\theta_4(2y)\theta_1(2z)\theta_1(2t)
\cr
&\qquad
=\theta_4(x+y+z+t)\theta_4(x+y-z-t)\theta_1(x-y-z+t)\theta_1(x-y+z-t)
\cr
&\qquad
+\theta_4(x+y+z-t)\theta_4(x+y-z+t)\theta_1(x-y+z+t)\theta_1(-x+y+z+t),
\lnlabel\EQDthetaid}
$$
we obtain
$$
\eqalignno{
H^{(i)}_m(u,u_0;v)
&=(-)^i\sum_n\e^{-2\epsilon{r-1\over r}(v-u+{1\over4})
-2\epsilon(v-u+{1\over2})m-\epsilon{r\over r-1}m^2
-\epsilon{r-1\over r}n^2+2\epsilon mn
+2\epsilon{r-1\over r}(v-u+{1\over2})n}
\cr
&\quad\times
{h_4(v-u_0-\half)\,h_4(v-u+\half-n)
\over h_1(v-u_0-\half)\,h_1(v-u-\half)}.
\lnlabel\EQDHsimp}
$$
From the summand we can extract an $n$-dependent factor
$$
h^{(i)}_{m,n}
=\e^{-\epsilon{r-1\over r}n^2+2\epsilon mn+2\epsilon{r-1\over r}wn}
\,\theta_4\!\left({w-n\over r};{\i\pi\over\epsilon r}\right)
=\sqrt{\epsilon r\over\pi}\,
\e^{-\epsilon{r-1\over r}n^2+2\epsilon mn+2\epsilon{r-1\over r}wn
-{\epsilon\over r}(w-n)^2}
\theta_2\!\left(\i{\epsilon\over\pi}(w-n);\i{\epsilon r\over\pi}\right)
$$
with $w=v-u+\half$. Expanding the last theta function in a series,
we obtain
$$
\sum_{n\in2\Z+m+i}h_{m,n}^{(i)}
=\sqrt{\epsilon r\over\pi}
\sum_{p\in\Z+\half(m+i)}\sum_{s\in\Z+\half}
\e^{-\epsilon rs^2-2\epsilon ws-4\epsilon p^2
+4\epsilon mp+4\epsilon wp+4\epsilon sp
-{\epsilon\over r}w^2}.
$$
Now we can split the sum into two terms: the first with even $s-m-\half$
and the second with even $s-m+\half$. Each term factors after change
of summation variables $(p,s)$ to $(k,l)$ so that $s=m\pm\half+2l$
and $p=k+l$. Each sum in $k$ and $l$ gives a theta function.
$$
\eqalignno{
H^{(i)}_m(u,u_0;v)
&=(-)^i\e^{\quarter\epsilon r-2\epsilon{r-1\over r}(v-u+\quarter)
+\epsilon{r-1\over r}(v-u+\half)^2}
{h_4(v-u_0-\half)\over h_1(v-u_0-\half)\,h_1(v-u-\half)}
\cr
&\quad\times
\biggl\{\e^{-{\epsilon\over r-1}\left(rm-{r-1\over2}\right)^2
-\epsilon\left(w+{1\over2}\right)^2}
\theta_3\!\left(\i{2\epsilon\over\pi}\left(rm-{r-1\over2}\right);
                        \i{4\epsilon(r-1)\over\pi}\right)
\theta_{3,2}\!\left(\i{2\epsilon\over\pi}\left(w+{1\over2}\right);
			\i{4\epsilon\over\pi}\right)
\cr
&\quad
+\e^{-{\epsilon\over r-1}\left(rm+{r-1\over2}\right)^2
-\epsilon\left(w-{1\over2}\right)^2}
\theta_3\!\left(\i{2\epsilon\over\pi}\left(rm+{r-1\over2}\right);
                        \i{4\epsilon(r-1)\over\pi}\right)
\theta_{3,2}\!\left(\i{2\epsilon\over\pi}\left(w-{1\over2}\right);
			\i{4\epsilon\over\pi}\right)\biggr\}.
}
$$
Here the subscript 3 or 2 depends on evenness of $m+i$.
Making the module conjugation transformation we get
$$
\eqalignno{
H^{(i)}_m(u,u_0;v)
&={(-1)^i\over4\sqrt{r-1}}\,
\e^{\quarter\epsilon r-2\epsilon{r-1\over r}(v-u+\quarter)
+\epsilon{r-1\over r}(v-u+\half)^2}
{h_4(v-u_0-\half)\over h_1(v-u_0-\half)\,h_1(v-u-\half)}
\cr
&\quad\times
\biggl\{
\theta_3\!\left({1\over2}{r\over r-1}m-{1\over4};
                          {\i\pi\over4\epsilon(r-1)}\right)
\theta_{3,4}\!\left({w\over2}+{1\over4};{\i\pi\over4\epsilon}\right)
\cr
&\quad
+\theta_3\!\left({1\over2}{r\over r-1}m+{1\over4};
                          {\i\pi\over4\epsilon(r-1)}\right)
\theta_{3,4}\!\left({w\over2}-{1\over4};{\i\pi\over4\epsilon}\right)\biggr\}.
\lnlabel\EQDHresult}
$$
Here the subscript 3 of the second and forth theta functions corresponds to
even $m+i$ and the index 4 to odd one. In principle, the above resummation
procedure is the same as that of the Appendix B.

Calculating $H^{(i)}_m-H^{(i)}_{-m}$ it is convenient to use the identity
$$
\textstyle
\theta_3(u+\quarter;\tau)-\theta_3(u-\quarter;\tau)
=-2\,\theta_1(2u;4\tau).
$$
Applying it twice, we get
$$
H^{(i)}_m-H^{(i)}_{-m}
=\e^{-2\epsilon{r-1\over r}(v-u+\quarter)+\epsilon{r-1\over r}(v-u+\half)^2}
{h_4(v-u_0-\half)\,\vartheta_1(v-u-\half)
\over h_1(v-u_0-\half)\,h_1(v-u-\half)}\,[m]'.
\eqlabel\EQDHdiffresult
$$
The factor $[m]'$ here will cancel the respective quantity
in the denominator in Eq.~(\EQstaggeredtrick). The $u$ depending
theta functions here will cancel theta functions in $G(v)$.

Gathering all results we obtain
$$
\eqalignno{
\langle\ve\rangle^{(i)}
&=\i(-1)^i{(x^2;x^4)_\infty\over[m]'}\,
\epsilon\int{dv\over\i\pi}\,
(H^{(i)}_m(u,u_0;v)-H^{(i)}_{-m}(u,u_0;v))\,G(v-u)
\cr
&=-(-1)^i{(x^2;x^2)_\infty^2\over(-x^2;x^2)_\infty^2}I,
\lnlabel\EQDcorri}
$$
where
$$
I=\epsilon\int_{C_0}{dv\over\i\pi}\,
{\theta_4\!\left({v-1/2\over r};{\i\pi\over\epsilon r}\right)
\over\theta_1\!\left({v-1/2\over r};{\i\pi\over\epsilon r}\right)}.
\eqlabel\EQDintegral
$$

To calculate the integral note that
$$
\theta_1(u+1)=-\theta_1(u),
\qquad
\theta_4(u+1)=\theta_4(u),
$$
and
$$
I=-\epsilon\int_{C_0}{dv\over\i\pi}\,
{\theta_4\!\left({v+r-1/2\over r};{\i\pi\over\epsilon r}\right)
\over\theta_1\!\left({v+r-1/2\over r};{\i\pi\over\epsilon r}\right)}
=-\epsilon\int_{C_r}{dv\over\i\pi}\,
{\theta_4\!\left({v-1/2\over r};{\i\pi\over\epsilon r}\right)
\over\theta_1\!\left({v-1/2\over r};{\i\pi\over\epsilon r}\right)}.
\eqlabel\EQDintegraltwo
$$
But the sum of (\EQDintegraltwo) and (\EQDintegral) is an
integral over a closed path enclosing the pole $v=1/2$. Hence
$$
I=-{\epsilon\over2}\oint{dv\over\i\pi}\,
{\theta_4\!\left({v-1/2\over r};{\i\pi\over\epsilon r}\right)
\over\theta_1\!\left({v-1/2\over r};{\i\pi\over\epsilon r}\right)}
=-\epsilon\Res_{v=1/2}{\theta_4\!\left({v-1/2\over r};{\i\pi\over\epsilon r}\right)
\over\theta_1\!\left({v-1/2\over r};{\i\pi\over\epsilon r}\right)}
=-\epsilon r{\theta_4(0;{\i\pi\over\epsilon r})
\over\theta'_1(0;{\i\pi\over\epsilon r})}.
$$
Using the multiplicative form of the theta functions we easily obtain
$$
I=-{(-x^{2r};x^{2r})_\infty^2\over(x^{2r};x^{2r})_\infty^2}.
\eqlabel\EQDintres
$$
Substituting (\EQDintres) into (\EQDcorri) we obtain the Baxter--Kelland
formula (\EQBK).

\bigskip\allowbreak\bigskip\immediate\closeout\rfile
\vbox{\secfont\noindent References\bigskip}\nobreak
\catcode`@=11\input refs.tmp\catcode`@=12\bigskip

\end